\documentclass[journal]{IEEEtran}
\usepackage{cite,color}

\ifCLASSINFOpdf
\else
\fi

\usepackage{amsmath,amssymb,amsfonts}

\usepackage{soul}

\usepackage{xcolor}

\usepackage{algorithmic}
\usepackage{algorithm}

\usepackage{array}

\usepackage{url}

\usepackage{float}
\usepackage{subfigure,multirow,bm,stfloats}
\usepackage{textcomp}
\usepackage{amsthm}
\newtheorem{myDef}{Definition}
\newtheorem{Pro}{Property}
\newtheorem{theorem}{Theorem}
\newtheorem{lemma}{Lemma}
\usepackage{graphicx}

\begin{document}
%
\title{Introducing Hypergraph Signal Processing: Theoretical Foundation and Practical Applications}
%
%
%

\author{Songyang~Zhang, Zhi~Ding,~\IEEEmembership{Fellow,~IEEE},
        and~Shuguang~Cui,~\IEEEmembership{Fellow,~IEEE}
        \thanks{© 20XX IEEE.  Personal use of this material is permitted.  Permission from IEEE must be obtained for all other uses, in any current or future media, including reprinting/republishing this material for advertising or promotional purposes, creating new collective works, for resale or redistribution to servers or lists, or reuse of any copyrighted component of this work in other works.}}
\maketitle

\begin{abstract}
Signal processing over graphs has recently attracted significant attentions
for dealing with structured data. Normal graphs, however, only model pairwise relationships 
between nodes and are not effective in representing and capturing some high-order 
relationships of data samples, which are common in many applications such as Internet of Things (IoT).
In this work, we propose a new framework of hypergraph signal processing (HGSP) 
based on tensor representation to generalize the traditional graph signal processing (GSP) 
to tackle high-order interactions. We introduce the core concepts of HGSP
and define the hypergraph Fourier space. 
We then study the spectrum properties of hypergraph Fourier transform 
and explain its connection to mainstream digital signal processing. 
We derive the novel hypergraph sampling theory and present the fundamentals
of hypergraph filter design based on the tensor framework. 
We present HGSP-based methods for several signal processing and data analysis 
applications. Our experimental results demonstrate
significant performance improvement using our HGSP framework over some traditional
signal processing solutions. 
\end{abstract}


%
\IEEEpeerreviewmaketitle

\section{Introduction}
\IEEEPARstart{G}{raph} theoretic tools
have recently found broad applications in data science 
owing to their power to model complex relationships in large 
structured datasets \cite{c2}. 
Big data, such as those representing social network interactions, Internet of Things (IoT) intelligence, biological 
connections, mobility and traffic patterns,
often exhibit complex structures that are challenging  
to many traditional tools \cite{c1}. Thankfully,
graphs provide good models for many such datasets as well as
the underlying complex relationships. 
A dataset with $N$ data points can be modeled as a graph of $N$ vertices, 
whose internal relationships can be captured by edges. 
For example, subscribing users in a communication or social network 
can be modeled as nodes while the physical interconnections or social 
relationships among users are represented as edges \cite{soc}.

Taking advantage of graph models in characterizing complex data structures, 
graph signal processing (GSP) has emerged as an exciting and promising new tool for 
processing large datasets with complex structures. 
A typical application of GSP is in
image processing, where image pixels are modeled as 
graph signals embedding in nodes while pairwise similarities between pixels 
are captured by edges \cite{c5}. 
By modeling images using graphs, tasks such as
image segmentation can take advantage of graph partition and GSP filters. 
Another example of GSP applications is in processing data 
from sensor networks \cite{c4}. 
Based on graph models directly built over network structures, 
a graph Fourier space could be defined according to the eigenspace of
a representing graph matrix such as the Laplacian or adjacency matrix
to facilitate data processing operations such as denoising \cite{den}, filter banks \cite{fb} 
and compression \cite{compre}.

\begin{figure*}[t]
	\centering
	\subfigure[Example of a trait and genes: the trait $t_1$ is triggered by three genes $v_1$, $v_2$, $v_3$ and the genes may also influence each other.]{
		\label{mw1}
				\includegraphics[height=2cm]{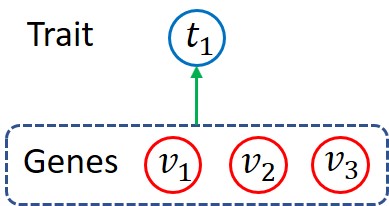}}
	\hspace{0.6in}
	\subfigure[Example of pairwise relationships: arrow links represent the influences from genes, whereas the potential interactions among genes cannot be captured.]{
		\label{mw2}
		\includegraphics[height=2cm]{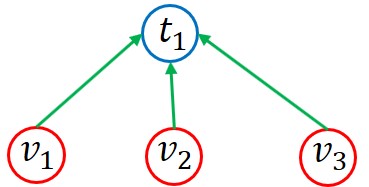}}
	\hspace{0.6in}
	\subfigure[Example of multi-lateral relationships: the solid circular line represents a quadrilateral relationship among four nodes, while the purple dash lines represent the potential inter-node interactions in this quadrilateral relationship.]{
		\label{mw3}
		\includegraphics[height=2cm]{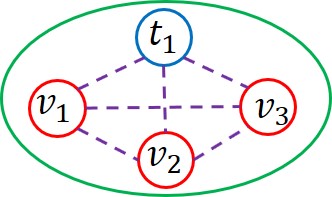}}
	\caption{Example of Multi-lateral Relationships.}
	\label{mw}
\end{figure*}

Despite many demonstrated successes, the GSP defined over normal graphs also exhibits
certain limitations. First, normal graphs cannot
capture high-dimensional interactions describing multi-lateral relationships among multiple nodes, which are critical for many practical applications.
Since each edge in a normal graph only models
the pairwise interactions between two nodes, the traditional GSP can 
only deal with the pairwise relationships defined by such edges. 
In reality, however, complex relationships may exist among a cluster of nodes, for which the use of pairwise links between every two nodes cannot capture their multi-lateral interactions \cite{c10}.
In biology, for example, a trait may be attributed to multiple interactive genes \cite{c11} shown in Fig. \ref{mw1}, such that a quadrilateral interaction is more informative and powerful here.
Another example is the social network with online social communities called folksonomies, 
where trilateral interactions occur among 
users, resources, and annotations \cite{c9,c30}. 
Second, a normal graph can only capture a typical single-tier relationship with matrix representation.
In complex systems and datasets, however, each node may have several traits such that there exist multiple tiers of interactions between two nodes.
In a cyber-physical system, for example, each node usually contains two components, i.e., the physical component and the cyber component, for which there exist two tiers of connections between a pair of nodes. Generally, such multi-tier relationships can be modeled as multi-layer networks, where each layer represents one tier of interactions \cite{mr1}. However, normal graphs cannot model the inter-layer interactions simply, and the corresponding matrix representations are unable to distinguish different tiers of relationships efficiently since they describe entries for all layers equivalently \cite{spec,mr5}.
Thus, the traditional GSP based on matrix analysis has far been unable to efficiently handle
such complex relationships. Clearly, 
there is a need for a more general graph model and graph signal processing
concept to remedy the aforementioned shortcomings faced with the traditional GSP. 

\begin{figure}[t]
	\centering
	\subfigure[Example of hypergraphs: the hyperedges are the overlaps covering nodes with different colors.]{
		\label{hg1}
		\includegraphics[height=3cm]{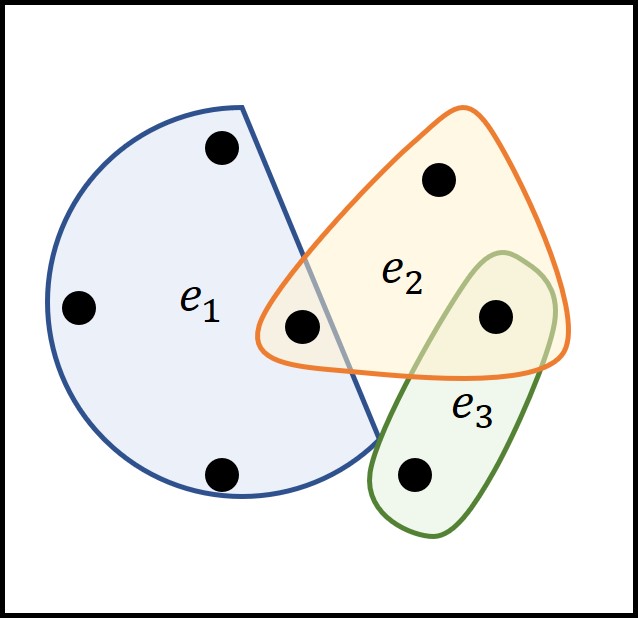}}
		\hspace{0.3in}
	\subfigure[A game dataset modeled by hypergraphs: each node is a specific game and each hyperedge is a catagory of games.]{
		\label{hg2}
		\includegraphics[height=3cm]{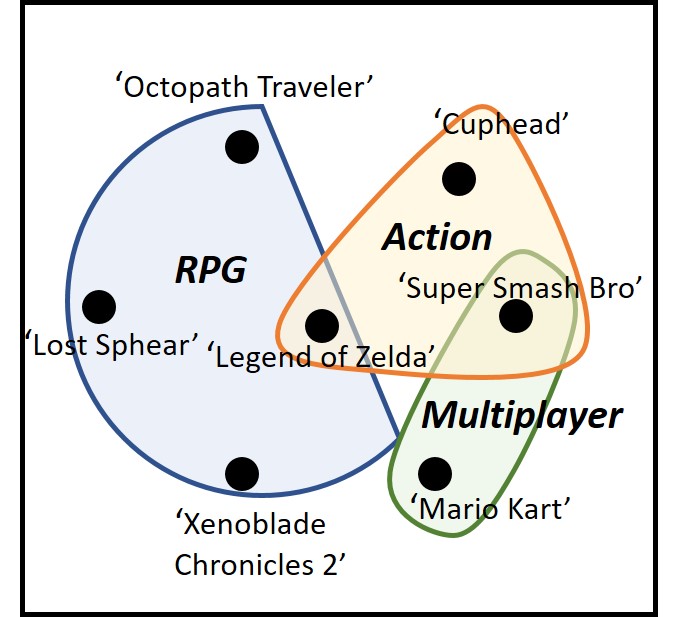}}
	\caption{Hypergraphs and Applications.}
	\label{hg}
\end{figure}

To find a more general model for complex data structures, we 
venture into the area of high-dimensional graphs known as hypergraphs. 
The hypergraph theory is playing an increasingly important role in graph theory and data analysis, 
especially for analyzing high-dimensional data structures
and interactions \cite{c12}. 
A hypergraph consists of nodes and hyperedges 
connecting more than two nodes \cite{c13}. As an example, 
Fig. \ref{hg1} shows
a hypergraph example with three hyperedges and seven nodes, whereas
Fig. \ref{hg2} provides
a corresponding dataset modeled by this hypergraph. 
Indeed, a normal graph is a special case of a hypergraph, 
where each hyperedge degrades to a simple edge that only involves exactly two nodes. 

Hypergraphs have found successes by generalizing
normal graphs in many applications, 
such as clustering \cite{c29}, classification \cite{GCN}, and prediction \cite{HP}. 
Moreover, a hypergraph is an alternative representation for a multi-layer network, and is useful when dealing with multi-tier relationships \cite{mr2,mr3}.
Thus, a hypergraph is a natural extension of a normal graph in modeling signals of 
high-degree interactions. Presently, however, the literature provides little
coverage on hypergraph signal processing (HGSP). 
The only known work \cite{c3} proposed a HGSP framework based on a 
special hypergraph called complexes. In this work \cite{c3}, 
hypergraph signals are associated with each hyperedge, but its
framework is limited to cell complexes, 
which cannot suitably model many real-world
datasets and applications.
Another shortcoming of the framework in \cite{c3} is the lack of detailed analysis and application examples
to demonstrate its practicability. In addition, the attempt in \cite{c3} to extend some key concepts 
from the traditional GSP simply fails due to the difference
in the basic setups between graph signals and hypergraph signals. 
In this work, we seek to establish a more general and practical
HGSP framework, capable of handling arbitrary hypergraphs 
and naturally extending the traditional GSP concepts to handle the high-dimensional 
interactions. We will also provide real application examples to validate the effectiveness of the proposed framework.

Compared with the traditional GSP, a generalized HGSP faces several technical challenges.
The first problem lies in the mathematical representation of hypergraphs.
Developing an algebraic representation of a hypergraph is the
foundation of HGSP. Currently there are two major approaches:  matrix-based \cite{c99} and 
tensor-based \cite{c14}.  The matrix-based method makes it hard to implement the
hypergraph signal shifting while the tensor-based method is difficult to be understood conceptually.
Another challenge is in defining signal shifting over the hyperedge.
Signal shifting is easy to be defined as propagation along the link direction 
of a simple edge connecting two nodes in a regular graph. 
However, each hyperedge in hypergraphs involves more than two nodes.
How to model signal interactions over a hyperedge requires careful considerations.
Other challenges include the definition and interpretation of hypergraph frequency.

To address the aforementioned challenges and generalize the traditional GSP 
into a more general hypergraph tool to capture high dimension interactions, 
we propose a novel tensor-based HGSP framework in this paper. 
The main contributions in this work can be summarized as follows.
Representing hypergraphs as tensors, we define a specific form of 
hypergraph signals and hypergraph signal shifting. 
We then provide an alternative definition of hypergraph Fourier space
based on the orthogonal CANDECOMP/PARAFAC (CP) tensor decomposition, together with the corresponding hypergraph Fourier transform. 
To better interpret the hypergraph Fourier space, we analyze the resulting hypergraph
frequency properties, including the concepts of frequency and bandlimited signals. 
Analogous to the traditional sampling theory, we derive the conditions and properties for
perfect signal recovery from samples in HGSP. We also provide the theoretical 
foundation for the HGSP filter designs. 
Beyond these, we provide several application examples of
the proposed HGSP framework: 
\begin{itemize}
\item[1)] We introduce a signal compression method based on the new sampling theory to 
show the effectiveness of HGSP in describing structured signals; 
\item[2)] We apply HGSP in spectral clustering to show 
how the HGSP spectrum space acts as a suitable spectrum for hypergraphs; 

\item[3)]  We introduce a HGSP method for binary classification problems to
demonstrate the practical application of HGSP in data analysis; 
\item[4)] We introduce a filtering approach for the denoising problem to 
further showcase the power of HGSP;
\item[5)] Finally, we suggest several potential applicable background for HGSP, including Internet of Things (IoT), social network and nature language processing.
\end{itemize} 
We compare the performance of HGSP-based methods with the traditional 
GSP-based methods and learning algorithms in all the above applications. All the features of HGSP make it an essential tool for IoT applications in the future.

We organize the rest of the paper as follows. 
Section \uppercase\expandafter{\romannumeral2} first 
summarizes the preliminaries of the traditional GSP,
tensors, and hypergraphs.
In Section \uppercase\expandafter{\romannumeral3}, we then introduce the core definitions of HGSP, 
including the hypergraph signal, the signal shifting and the hypergraph Fourier space, followed by the frequency 
interpretation and decription of existing works in Section \uppercase\expandafter{\romannumeral4}.
We present some useful HGSP-based results such as the sampling theory 
and filter design in Section \uppercase\expandafter{\romannumeral5}. With the proposed HGSP framework, we provide several 
potential applications of HGSP and demonstrate its effectiveness 
in Section \uppercase\expandafter{\romannumeral6}, before 
presenting the final conclusions in
Section \uppercase\expandafter{\romannumeral7}.

\section{Preliminaries}

\subsection{Overview of Graph Signal Processing}
GSP is a recent tool used to analyze signals according to the graph models. 
Here, we briefly review the key relevant concepts of the traditional GSP \cite{c1,c2}. 

A dataset with $N$ data points can be modeled as a normal 
graph $\mathcal{G}(\mathcal{V,E})$ consisting of a set of $N$ nodes $\mathcal{V}=\{\mathbf{v}_1,\cdots,\mathbf{v}_N\}$ and a set of 
edges $\mathcal{E}$. Each node of the graph $\mathcal{G}$ 
is a data point, whereas the edges describe the pairwise interactions between nodes. 
A graph signal represents the data associated with a node. 
For a graph with $N$ nodes, there are $N$ graph signals, which are defined as a signal vector
$
\mathbf{s}=[s_1\quad s_2\quad...\quad s_N]^{\mathrm{T}} \in \mathbb{R}^{N}.
$

Usually, such a graph could be either described by an adjacency matrix $\mathbf{A_M}\in\mathbb{R}^{N\times N}$ where each entry indicates a pairwise link
(or an edge), or by a Laplacian matrix $\mathbf{L_M=D_M-A_M}$ where $\mathbf{D_M}\in \mathbb{R}^{N\times N}$ is the diagonal matrix of degrees. 
Both the Laplacian matrix and the adjacency matrix can fully
represent the graph structure. For convenience, we use a general matrix $\mathbf{F_M}\in \mathbb{R}^{N\times N}$ to represent either of them. 
Note that, since the adjacency matrix is eligible in both directed and undirected graph, it is more common in the GSP literatures. Thus, the generalized GSP is based on the adjacency matrix \cite{c1} and the representing matrix refers to the adjacency matrix in this paper unless specified otherwise.

\begin{figure}[t]
	\centering
	\includegraphics[width=2in]{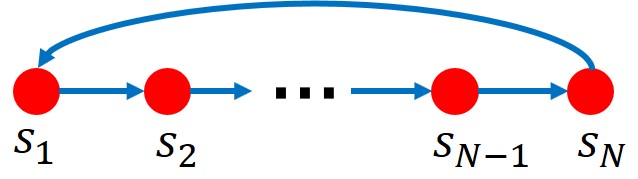}
	\caption{The Signal Shifting over Cyclic Graph.}
	\label{CIRC}
\end{figure}

With the graph representation $\mathbf{F_M}$ and the signal vector $\mathbf{s}$, the graph shifting is defined as
\begin{equation}\label{shift}
\mathbf{s'=F_Ms}.
\end{equation}
Here, the matrix $\mathbf{F_M}$ could be interpreted as a graph filter whose functionality is to shift the signals along link directions. Taking the cyclic graph 
shown in Fig. \ref{CIRC} as an example, its adjacency matrix 
is a shifting matrix
\begin{align}
	\mathbf{F_M}=
	\begin{bmatrix}
	0&0&\cdots&0&1\\
	1&0&\cdots&0&0\\
	\vdots&\ddots&\ddots&\ddots&\vdots\\
		0&0&\ddots&0&0\\
    0&0&\cdots&1&0
	\end{bmatrix}.
\end{align} 
Typically, the shifted signal over the cyclic graph is calculated as $\mathbf{s'=F_Ms}=[s_{N}\quad s_1\quad\cdots\quad s_{N-1}]^{\mathrm{T}}$, which shifts the signal at each node to its next node.

The graph spectrum space, also called the graph Fourier space, is defined based on the eigenspace of $\mathbf{F_M}$. Assume that the eigen-decomposition of $\mathbf{F_M}$ is
\begin{equation}\label{13}
\mathbf{F_M=V_M^{-1}\Lambda V_M}.
\end{equation}
The frequency components are defined by the eigenvectors of $\mathbf{F_M}$ and the frequencies are defined with respect to eigenvalues. The corresponding graph Fourier transform is defined as
\begin{equation}\label{44}
\mathbf{\hat{s}}=  \mathbf{V_Ms}.
\end{equation}

With the definition of the graph Fourier space, the traditional signal processing and learning tasks, such as denoising \cite{c22} and classification \cite{fl2}, could be solved within the GSP framework. More details about the specific topics of GSP, such as the frequency analysis, filter design, and spectrum representation have been discussed in \cite{c4,GSP1,gsp0}.

\subsection{Introduction of Hypergraph}
We begin with the definition of hypergraph and its possible representations.

\begin{myDef}[Hypergraph]
	A general hypergraph $\mathcal{H}$ is a pair $\mathcal{H}=(\mathcal{V,E})$, where $\mathcal{V}=\{\mathbf{v}_1,...,\mathbf{v}_N\}$ is a set of elements called vertices and $\mathcal{E}=\{\mathbf{e}_1,...,\mathbf{e}_K\}$ is a set of non-empty multi-element subsets of $\mathcal{V}$ called hyperedges. Let $M=\max\{|\mathbf{e}_i|:\mathbf{e}_i\in\mathcal{E}\}$ be the maximum cardinality of hyperedges, shorted as $m.c.e(\mathcal{H})$ of $\mathcal{H}$.
\end{myDef}

\begin{figure}[htbp]
	\centering
	\includegraphics[width=2in]{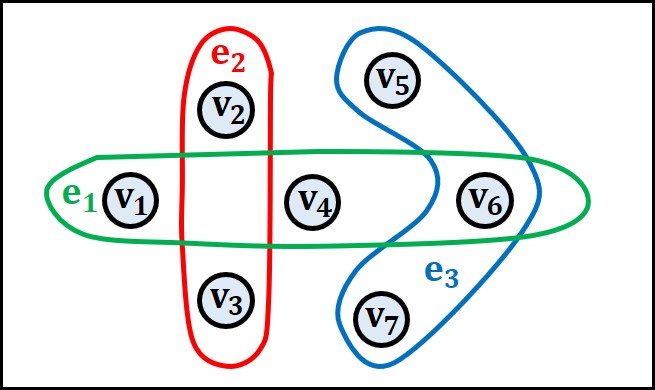}
	\caption{A hypergraph $\mathcal{H}$ with 7 nodes, 3 hyperedges and $m.c.e(\mathcal{H})=3$, where $\mathcal{V}=\{\mathbf{v}_1,\cdots, \mathbf{v}_7\}$ and $\mathcal{E}=\{\mathbf{e}_1, \mathbf{e}_2, \mathbf{e}_3\}$. Three hyperedges are $\mathbf{e}_1=\{\mathbf{v}_1, \mathbf{v}_4, \mathbf{v}_6\}$, $\mathbf{e}_2=\{\mathbf{v}_2, \mathbf{v}_3\}$ and $\mathbf{e}_3=\{\mathbf{v}_5, \mathbf{v}_6, \mathbf{v}_7\}$.}
	\label{hyper}
\end{figure}

In a general hypergraph $\mathcal{H}$, different hyperedges may contain 
different numbers of nodes. The $m.c.e(\mathcal{H})$ denotes the 
number of vertices in the largest hyperedge. An example of a hypergraph with 
$7$ nodes, $3$ hyperedges and $m.c.e=3$ is shown in Fig. \ref{hyper}.

From the definition, we see that a normal graph is a special case of a hypergraph 
if $M=2$. The hypergraph is a natural extension of the normal graph to represent high-dimensional
 interactions. To represent a hypergraph mathematically, there are two 
 major methods based on matrix and
tensor respectively. In the matrix-based method, a hypergraph is represented by a 
matrix $\mathbf{G}\in \mathbb{R}^{N\times E}$ where $E$ equals the 
number of hyperedges. 
The rows of the matrix represent the nodes, and the columns represent the hyperedges \cite{c13}. 
Thus, each element in the matrix indicates whether the corresponding node 
is involved in the particular hyperedge. 
Although such a matrix-based representation is simple in formation, 
it is hard to define and implement signal processing directly as in GSP 
by 
using the matrix $\mathbf{G}$. 
Unlike the matrix-based method, tensor has better flexibility
in describing the structures of the high-dimensional graphs \cite{luo}. 
More specifically, tensor can be viewed as an extension 
of matrix into high-dimensional domains. 
The adjacency tensor, which indicates whether nodes are connected, 
is a natural hypergraph counterpart to the adjacency matrix in the normal 
graph theory \cite{adt}. 
Thus, we prefer to represent the hypergraphs using tensors. In Section III-A, 
we will provide more details on how to represent the hypergraphs and signals 
in tensor forms.

\subsection{Tensor Basics}
Before we introduce our tensor-based HGSP framework, let us introduce some tensor 
basics to be used later. Tensors can effectively represent
high-dimensional graphs \cite{mr5}. Generally speaking, tensors can be interpreted 
as multi-dimensional arrays. The order of a tensor is the number of 
indices needed to label a component of that array\cite{c15}. 
For example, a third-order tensor has three indices.
In fact, scalars, vectors and matrices 
are all special cases of tensors: a scalar is a zeroth-order tensor; 
a vector is a first-order tensor; a matrix is a second-order tensor; and 
an $M$-dimensional array is an $M$th-order tensor \cite{spec}. Generalizing a 2-D matrix, we represent the entry at the position $(i_1,i_2,\cdots,i_M)$ of an $M$th-order tensor $\mathbf{T}\in\mathbb{R}^{I_1\times I_2\times\cdots\times I_M}$ by $t_{i_1i_2\cdots i_M}$ in the rest of the paper.

Below are some useful definitions and operations of tensor related to the proposed HGSP framework.
\subsubsection{Symmetric and Diagonal Tensors}
\begin{itemize}
	\item A tensor is \textit{super-symmetric} if its entries are invariant under any permutation of their indices \cite{spt}. For example, a third-order $\mathbf{T}\in \mathbb{R}^{I\times I \times I}$ is super-symmetric if its entries $t_{ijk}$'s 
	satisfy
	\begin{equation}
	t_{ijk}=t_{jik}=t_{kij}=t_{kji}=t_{jik}=t_{jki}\quad i,j,k=1,\cdots,I.
	\end{equation}
	Analysis of super-symmetric tensors, which is shown to be bijectively related
	to homogeneous polynomials, could be found in \cite{sym1,sym2}.
	\item A tensor $\mathbf{T}\in\mathbb{R}^{I_1\times I_2\cdots\times I_N}$ is \textit{super-diagonal} if its entries $t_{i_1i_2\cdots i_N}\neq 0$ only if $i_1=i_2=\cdots=i_N$. For example, a third-order $\mathbf{T}\in \mathbb{R}^{I\times I \times I}$ is super-diagonal if its entries $t_{iii}\neq 0$ for $i=1,2,\cdots,I$, 
	while all other entries are zero.
\end{itemize}

\subsubsection{Tensor Operations}
Tensor analysis is developed based on tensor operations. Some tensor operations
are commonly used in our HGSP framework \cite{katri,hada,n-mode}.
\begin{itemize}
	\item The \textit{tensor outer product} between an $P$th-order tensor $\mathbf{U}\in \mathbb{R}^{I_1\times I_2\times ...\times I_P }$ with entries $u_{i_1 ... i_P}$ and an $Q$th-order tensor $\mathbf{V}\in \mathbb{R}^{J_1\times J_2\times ...\times J_Q }$ with entries $v_{j_1 ... j_Q}$ is denoted by $\mathbf{W}=\mathbf{U} \circ \mathbf{V}$. The result $\mathbf{W}\in \mathbb{R}^{I_1\times I_2\times ...\times I_P \times J_1 \times J_2 \times ... \times J_Q}$ is an $(P+Q)$-th order tensor, whose entries are calculated by
	\begin{equation}
	w_{i_1 ... i_P j_1 ... j_Q}= u_{i_1 ... i_P} \cdot v_{j_1 ... j_Q}.
	\end{equation}
The major use of the tensor outer product is to construct a higher order tensor with several lower order tensors.
	For example, the tensor outer product between vectors $\mathbf{a}\in\mathbb{R}^{M}$ and $\mathbf{b}\in\mathbb{R}^{N}$ is denoted by
	\begin{equation}
		\mathbf{T=a\circ b},
	\end{equation}
	where the result $\mathbf{T}$ is a matrix in $\mathbb{R}^{M\times N}$ with entries $t_{ij}=a_i\cdot b_j$ for $i=1,2,\cdots,M$ and $j=1,2,\cdots,N$. Now, we introduce one more vector $\mathbf{c}\in\mathbb{R}^{Q}$, 
	where
	\begin{equation}
		\mathbf{S=a\circ b\circ c=T\circ c}.
	\end{equation}
	Here, the result $\mathbf{S}$ is a third-order tensor with entries $s_{ijk}=a_i\cdot b_j \cdot c_k=t_{ij}\cdot c_k$ for $i=1,2,\cdots, M$, $j=1,2,\cdots, N$ and $k=1,2,\cdots, Q$.
	\item The \textit{n-mode product} between a tensor $\mathbf{U}\in \mathbb{R}^{I_1\times I_2 \times \cdots \times I_P}$ and a matrix $\mathbf{V\in \mathbb{R}^{J\times I_n}}$ is denoted by $\mathbf{W}=\mathbf{U\times_n V}\in \mathbb{R}^{I_1\times I_2\times \cdots\times I_{n-1}\times J \times I_{n+1} \times \cdots \times I_P}$. Each element in $\mathbf{W}$ is defined as
	\begin{equation}\label{nmode}
	w_{i_1 i_2 \cdots i_{n-1} j i_{n+1} \cdots i_P}=\sum_{i_n=1}^{I_n}u_{i_1\cdots i_P}v_{j i_n},
	\end{equation}
	where the main function is to adjust the dimension of a specific order. For example, in Eq. (\ref{nmode}), the dimension of the $n$th order of $\mathbf{U}$ is changed from $I_n$ to $J$.	
	\item The \textit{Kronecker product} of matrices $\mathbf{U}\in \mathbb{R}^{I\times J}$ and $\mathbf{V}\in \mathbb{R}^{P\times Q}$ is defined as
	\begin{subequations}
	\begin{align}
	\mathbf{U\otimes V}&=
	\begin{bmatrix}
	u_{11}\mathbf{V}& u_{12}\mathbf{V}& \cdots&u_{1J}\mathbf{V}\\
	u_{21}\mathbf{V}& u_{22}\mathbf{V}& \cdots&u_{2J}\mathbf{V}\\
	\vdots&\vdots& \ddots&\vdots\\
	u_{I1}\mathbf{V}& u_{I2}\mathbf{V}& \cdots&u_{IJ}\mathbf{V}
	\end{bmatrix}
	\end{align}
	\end{subequations}
	to generate an $IP\times JQ$ matrix.

	\item The \textit{Khatri-Rao product} between 
	$\mathbf{U}\in\mathbb{R}^{I\times K}$ and $\mathbf{V}\in\mathbb{R}^{J\times K}$ is defined as 
	\begin{equation}
	\mathbf{U\odot V}=[\mathbf{u}_1\otimes \mathbf{v}_1\quad \mathbf{u}_2\otimes \mathbf{v}_2\quad\cdots\quad \mathbf{u}_K\otimes \mathbf{v}_K].
	\end{equation}

	\item	The \textit{Hadamard product} between $\mathbf{U}\in\mathbb{R}^{P\times Q}$ and $\mathbf{V}\in\mathbb{R}^{P\times Q}$ is defined as 
	\begin{align}
	\mathbf{U*V}=
	\begin{bmatrix}
	u_{11}v_{11}& u_{12}v_{12}& \cdots&u_{1Q}v_{1Q}\\
	u_{21}v_{21}& u_{22}v_{22}& \cdots&u_{2Q}v_{2Q}\\
	\vdots&\vdots& \ddots&\vdots\\
	u_{P1}v_{P1}& u_{P2}v_{P2}& \cdots&u_{PQ}v_{PQ}
	\end{bmatrix}.
	\end{align}	
\end{itemize}

\subsubsection{Tensor Decomposition}
\begin{figure}[t]
	\centering
	\includegraphics[width=2.8in]{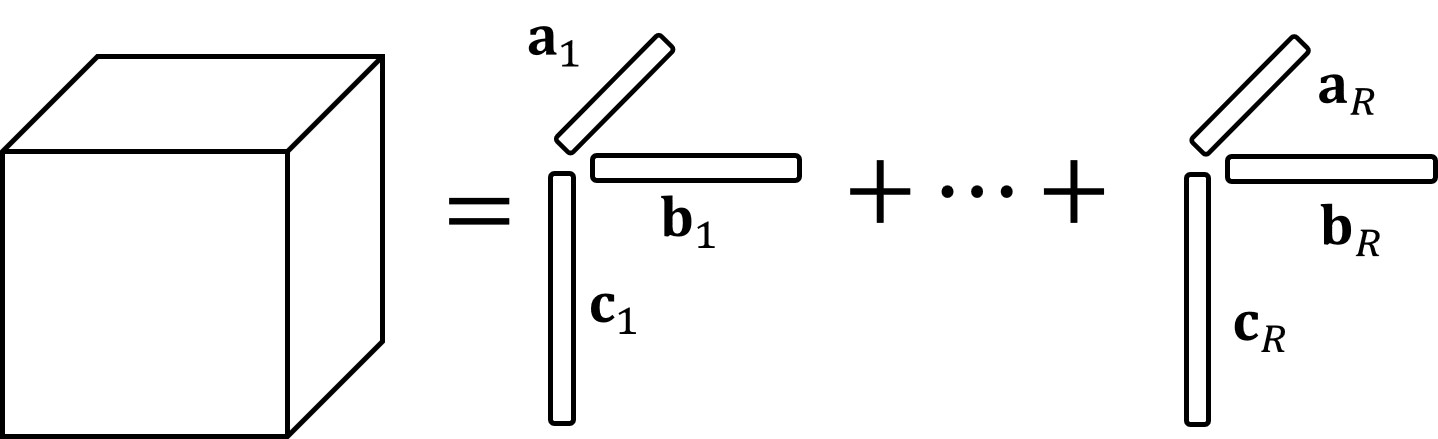}
	\caption{CP Decomposition of a Third-order Tensor.}
	\label{TD}
\end{figure}

Similar to the eigen-decomposition for matrix, tensor decomposition 
analyzes tensors via factorization. The CANDECOMP/PARAFAC (CP)
decomposition is a widely used method, which factorizes a tensor 
into a sum of component rank-one 
tensors \cite{c15,decomp}. For example, a third order tensor $\mathbf{T}\in\mathbb{R}^{I\times J\times K}$ 
is decomposed into
\begin{equation}
\mathbf{T}=\sum_{r=1}^{R}\mathbf{a}_r\circ \mathbf{b}_r\circ \mathbf{c}_r,
\end{equation}
where $\mathbf{a}_r\in \mathbb{R}^I$, $\mathbf{b}_r\in \mathbb{R}^J$, $\mathbf{c}_r\in \mathbb{R}^K$ and $R$ is a positive integer known as
rank, which leads to the smallest number of rank-one tensors in the decomposition. The process of CP decomposition for a third-order tensor is illustrated in Fig. \ref{TD}.

There are several extensions and alternatives of the CP decomposition. 
For example, the orthogonal-CP decomposition \cite{c17} decomposes 
the tensor 
using an orthogonal basis. For an $M$-th order $N$-dimension tensor $\mathbf{T}\in\mathbb{R}^{\underbrace{\scriptstyle{N\times 
			N\times...\times N}}_{\text{M times}}}$, it can 
be decomposed by the orthogonal-CP decomposition as
\begin{equation}
\mathbf{T}\approx\sum_{r=1}^{R}\lambda_r\cdot\mathbf{a}_r^{(1)}\circ...\circ \mathbf{a}_r^{(M)},
\end{equation}
where $\lambda_r\geq 0$ and the orthogonal basis is $\mathbf{a}_r^{(i)}\in\mathbb{R}^N$ for 
$1\leq i\leq M$. More specifically, the
orthogonal-CP decomposition 
has a similar form to the eigen-decomposition when $M=2$ and $\mathbf{T}$ 
is super-symmetric.

\subsubsection{Tensor Spectrum} The eigenvalues and spectral space of tensors are 
significant topics in tensor algebra. The research of 
tensor spectrum has achieved great progress in recent years. It will take a large volume to
cover all the properties of the tensor spectrum. Here, we just list some 
helpful and relevant literatures. In particular,
Lim and the others developed theories of eigenvalues, eigenvectors, singular values,
and singular vectors for tensors based on a constrained variational
approach such as the Rayleigh quotient \cite{tspec2}. 
Qi and the others in \cite{c16,tspec1} presented a more complete discussion
of tensor eigenvalues by defining two forms of tensor eigenvalues, 
i.e., the E-eigenvalue and the H-eigenvalue. Chang and the others \cite{spt} further extended the work of \cite{c16,tspec1}. 
Other works including \cite{tspec3,tspec4} 
further developed the theory of tensor spectrum.

\section{Definitions for Hypergraph Signal Processing}
In this section, we introduce the core definitions used in our HGSP framework. 

\subsection{Algebraic Representation of Hypergraphs}
The traditional GSP mainly relies on the representing matrix of a graph. 
Thus, an effective algebraic representation
is also helpful in developing a novel HGSP framework. 
As we mentioned in Section II-C, tensor is an intuitive representation 
for high-dimensional graphs. 
In this section, we introduce the algebraic representation of hypergraphs 
based on tensors.
\begin{figure*}[t]
	\centering
	\subfigure[A $3$-uniform hypergraph $\mathcal{H}$ with 7 nodes, 3 hyperedges and $m.c.e(\mathcal{H})=3$, where $\mathcal{V}=\{\mathbf{v}_1,\cdots, \mathbf{v}_7\}$ and $\mathcal{E}=\{\mathbf{e}_1, \mathbf{e}_2, \mathbf{e}_3\}$. Three hyperedges are $\mathbf{e}_1=\{\mathbf{v}_1, \mathbf{v}_4, \mathbf{v}_6\}$, $\mathbf{e}_2=\{\mathbf{v}_2, \mathbf{v}_3,\mathbf{v}_7\}$ and $\mathbf{e}_3=\{\mathbf{v}_5, \mathbf{v}_6, \mathbf{v}_7\}$.]{
		\label{hyper3}
		\includegraphics[height=3.5cm]{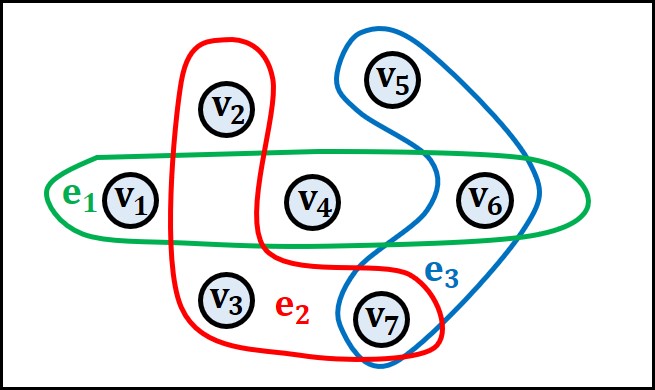}}
	\hspace{1in}
	\subfigure[A general hypergraph $\mathcal{H}$ with 7 nodes, 3 hyperedges and $m.c.e(\mathcal{H})=3$, where $\mathcal{V}=\{\mathbf{v}_1,\cdots, \mathbf{v}_7\}$ and $\mathcal{E}=\{\mathbf{e}_1, \mathbf{e}_2, \mathbf{e}_3\}$. Three hyperedges are $\mathbf{e}_1=\{\mathbf{v}_1, \mathbf{v}_4, \mathbf{v}_6\}$, $\mathbf{e}_2=\{\mathbf{v}_2, \mathbf{v}_3\}$ and $\mathbf{e}_3=\{\mathbf{v}_5, \mathbf{v}_6, \mathbf{v}_7\}$.]{
		\label{hyperg}
		\includegraphics[height=3.5cm]{Hypergraph.jpg}}
	\caption{Examples of Hypergraphs}
	\label{ex}
\end{figure*}

Similar to the adjacency matrix whose 2-D entries indicate whether and
how two nodes are pairwise connected by a simple edge, we adopt
an adjacency tensor whose entries indicate whether and how 
corresponding subsets of $M$ nodes are connected by hyperedges to describe hypergraphs\cite{c14}. 

\begin{myDef}[Adjacency tensor]
	A hypergraph $\mathcal{H}=(\mathcal{V,E})$ with $N$ nodes and $m.c.e(\mathcal{H})=M$ can be represented by an $M$th-order $N$-dimension adjacency tensor $\mathbf{A}\in\mathbb{R}^{\underbrace{\scriptstyle{N\times N\times\cdots\times N}}_{\text{M times}}}$ defined as
	\begin{equation}\label{ad}
	\mathbf{A}=(a_{i_1i_2\cdots i_M}),\quad 1\leq i_1,i_2,\cdots,i_M\leq N.
	\end{equation}
	
	Suppose that $\mathbf{e}_l=\{\mathbf{v}_{l1},\mathbf{v}_{l2},\cdots,\mathbf{v}_{lc}\}\in\mathcal{E}$ is a hyperedge in $\mathcal{H}$ with the number of vertices $c\leq M$. Then, $\mathbf{e}_l$ is represented by all the elements $a_{p_1\cdots p_M}$'s in $\mathbf{A}$, where a subset of $c$ indices from $\{p_1,p_2,\cdots,p_M\}$ are exactly the same as $\{l_1,l_2,\cdots,l_c\}$ and the other $M-c$ indices are picked from $\{l_1,l_2,\cdots,l_c\}$ randomly. More specifically, these elements $a_{p_1\cdots p_M}$'s describing $\mathbf{e}_l$ are calculated as
	\begin{equation}\label{adw}
	{a_{p_1\cdots p_M}={c}
\left(\sum_{{k_1,k_2,\cdots,k_c\geq 1, \sum_{i=1}^{c} k_i=M}} {\frac{M!}{k_1!k_2!...k_c!}}\right)^{-1}}.
	\end{equation}
	Meanwhile, the entries, which do not correspond to any hyperedge $\mathbf{e}\in\mathcal{E}$, are zeros.

\end{myDef}

Note that Eq. (\ref{adw}) enumerates all the possible combinations of $c$ positive integers $\{k_1,\cdots,k_c\}$, whose summation satisfies $\sum_{i=1}^{c} k_i=M$. Obviously, when the hypergraph degrades to the normal graph with $c=M=2$, the weights of edges are calculated as one, i.e., $a_{ij}=a_{ji}=1$ for an edge $\mathbf{e}=(i,j)\in\mathcal{E}$. Then, the adjacency tensor is the same as the adjacency matrix. To understand the physical meaning of the adjacency tensor and its weight, 
we start with the $M$-uniform hypergraph with $N$ nodes, where each 
hyperedge has exactly $M$ nodes\cite{kuni}. Since each hyperedge 
has an equal number of nodes, all hyperedges follow a consistent form 
to describe an $M$-lateral relationship with $m.c.e=M$. 
Obviously, such $M$-lateral relationships can be represented by 
an $M$th-order tensor $\mathbf{A}$, 
where the entry $a_{i_1i_2\cdots i_M}$ indicates whether the nodes $\mathbf{v}_{i_1},\mathbf{v}_{i_2},\cdots, \mathbf{v}_{i_M}$ 
are in the same hyperedge, i.e., whether a hyperedge $\mathbf{e}=\{\mathbf{v}_{i_1},\mathbf{v}_{i_2},\cdots, \mathbf{v}_{i_M}\}$ exists.
If the weight is nonzero, the hyperedge exists; otherwise, the hyperedge does not exist. 
Taking the $3$-uniform hypergraph in Fig. \ref{hyper3} as an example, 
the hyperedge $\mathbf{e}_1$ is characterized by $a_{146}=a_{164}=a_{461}=a_{416}=a_{614}=a_{641}\neq 0$, 
the hyperedge $\mathbf{e}_2$ is characterized by
$a_{237}=a_{327}=a_{732}=a_{723}=a_{273}=a_{372}\neq 0$,
and $\mathbf{e}_3$ is represented by 
$a_{567}=a_{576}=a_{657}=a_{675}=a_{756}=a_{765}\neq 0$. 
All other entries in $\mathbf{A}$ are zero.
Note that, all the hyperedges in an $M$-uniform hypergraph has the same weight. Different hyperedges are distinguished by the indices of the entries.
More specifically, similarly as $a_{ij}$ in the adjancency matrix implies the connection direction from node $\mathbf{v}_j$ to node $\mathbf{v}_i$ in GSP, an entry $a_{i_1i_2\cdots i_M}$ characterizes one direction of the hyperedge $\mathbf{e}=\{\mathbf{v}_{i_1},\mathbf{v}_{i_2},\cdots, \mathbf{v}_{i_M}\}$ with node $\mathbf{v}_{i_M}$ as the source and node $\mathbf{v}_{i_1}$ as the destination.

\begin{figure}[t]
\centering
\includegraphics[width=3in]{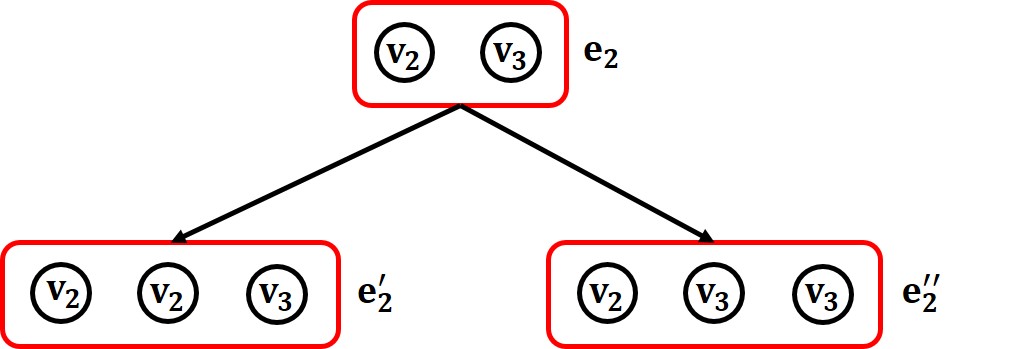}
\caption{Interpretation of Generalizing $\mathbf{e}_2$ to Hyperedges with $M=3$.}
	\label{norm3}
\end{figure}

However, for a general hypergraph, different hyperedges may contain 
different numbers of nodes. For example, in the hypergraph of
Fig. \ref{hyperg}, the hyperedge $\mathbf{e}_2$ only contains 
two nodes. How to represent the hyperedges with the number of nodes 
below $m.c.e=M$ may become an issue. 
To represent such a hyperedge $\mathbf{e}_l=\{\mathbf{v}_{l_1},
\mathbf{v}_{l_2},...,\mathbf{v}_{l_c}\}\in\mathcal{E}$ 
with the number of vertices $c< M$ in an $M$th-order tensor, 
we can use entries $a_{i_1i_2\cdots i_M}$, where a subset of $c$ indices are the same as $\{l_1,\cdots,l_c\}$ (possibly a different order) and the other $M-c$ indices are picked from $\{l_1,\cdots,l_c\}$ randomly. 
This process can be interpreted as generlaizing the hyperedge with $c$ nodes to a hyperedge with $M$ nodes by 
duplicating $M-c$ nodes from the set $\{\mathbf{v}_{l_1}, \cdots,\mathbf{v}_{l_c}\}$ randomly
with possible repetitions. 
For example,  
the hyperedge $\mathbf{e}_2=\{\mathbf{v}_2,\mathbf{v}_3\}$ in Fig. \ref{hyperg} can be represented by the entries $a_{233}=a_{323}=a_{332}=a_{322}=a_{223}=a_{232}$ 
in the third-order tensor $\mathbf{A}$, which could be interpreted as generalizing the original hyperedge with $c=2$
to hyperedges with $M=3$ nodes as Fig. \ref{norm3}.
We can use Eq. (\ref{adw}) as a generalization coefficient of each hyperedge 
with respect to permutation and combination \cite{c14}. More specifically, 
for the adjacency tensor of the hypergraph in Fig. \ref{hyperg}, 
the entries are calculated as $a_{146}=a_{164}=a_{461}=a_{416}=a_{614}=a_{641}=a_{567}=a_{576}=a_{657}=a_{675}=a_{756}=a_{765}=\frac{1}{2}$, $a_{233}=a_{323}=a_{332}=a_{322}=a_{223}=a_{232}=\frac{1}{3}$, 
where the remaining entries are set to zeros. Note that, the weight is smaller if the original hyperedge has fewer nodes in Fig. \ref{hyperg}. More generally, based on the definition of adjacency tensor and 
Eq. (\ref{adw}), we can easily obtain the following property regarding the hyperedge weight.
\begin{Pro}
Given two hyperedges $\mathbf{e}_i=\{\mathbf{v}_1,\cdots,\mathbf{v}_I\}$ and $\mathbf{e}_j=\{\mathbf{v}_1,\cdots,\mathbf{v}_J\}$, the edgeweight $w(\mathbf{e}_i)$ of $\mathbf{e}_i$ is different from the edgeweight $w(\mathbf{e}_j)$ of $\mathbf{e}_j$ in the adjacency tensor $\mathbf{A}$, i.e., $w(\mathbf{e}_i)\neq w(\mathbf{e}_j)$, if $I\neq J$. Moreover, $w(\mathbf{e}_i)=w(\mathbf{e}_j)$ iff $I=J$.
\end{Pro}
This property can help identify the length of each hyperedge based on the weights in the adjacency tensor. Moreover, the edgeweights of two hyperedges with the same number of nodes are the same. Different hyperedges with the same number of nodes are distinguished by their indices of entries in an adjacency tensor.


The degree $d(\mathbf{v}_i)$, of a vertex $\mathbf{v}_i\in \mathcal{V}$, is the number of hyperedges containing $\mathbf{v}_i$, i.e.,
\begin{equation}
d(\mathbf{v}_i)=\sum_{j_1, j_2\cdots, j_{M-1}=1}^N a_{ij_1j_2\cdots j_{M-1}}.
\end{equation}

Then, the Laplacian tensor of the hypergraph $\mathcal{H}$ is defined as follows \cite{c14}.

\begin{myDef}[Laplacian tensor]
	Given a hypergraph $\mathcal{H=(V,E)}$ with $N$ nodes and $m.c.e(\mathcal{H})=M$, the Laplacian tensor is defined as
	\begin{equation}
	\mathbf{L=D-A} \in \mathbb{R}^{\underbrace{\scriptstyle{N\times N\times...\times N}}_{\text{M times}}}
	\end{equation}
which is an $M$th-order $N$-dimension tensor. 
Here, $\mathbf{D}=(d_{i_1i_2\cdots i_M})$ is also
an $M$th-order $N$-dimension super-diagonal tensor with nonzero elements of $d_{\underbrace{\scriptstyle{ii\cdots i}}_{\text{M times}}}=d(\mathbf{v}_i)$.
\end{myDef}

We see that both the adjacency and Laplacian tensors of a hypergraph $\mathcal{H}$ are super-symmetric. Moreover, when $m.c.e(\mathcal{H})=2$, they have similar forms to the adjacency and Laplacian matrices of undirected graphs respectively. Similar to
GSP, we use an $M$th-order $N$-dimension tensor $\mathbf{F}$ as a 
general representation of a given hypergraph $\mathcal{H}$ 
for convenience. As the adjacency tensor is more general,
the representing tensor $\mathbf{F}$ refers to the adjacency tensor in this paper 
unless specified otherwise.

\subsection{Hypergraph Signal and Signal Shifting} 
Based on the tensor representation of hypergraphs, we now provide definitions 
for the hypergraph signal. In the traditional GSP, each signal element is related to one node in the graph. Thus, 
the graph signal in GSP is defined as an $N$-length vector if there are $N$ nodes in the graph. 
Recall that the representing matrix of a normal graph can be 
treated as a graph filter, for which the 
basic form of the filtered signal is defined in Eq. (\ref{shift}). 
Thus, we could extend the definitions of the graph signal and signal shifting 
from the traditional GSP to HGSP based on the tensor-based filter implementation.

\begin{figure}[t]
	\centering
	\includegraphics[width=3in]{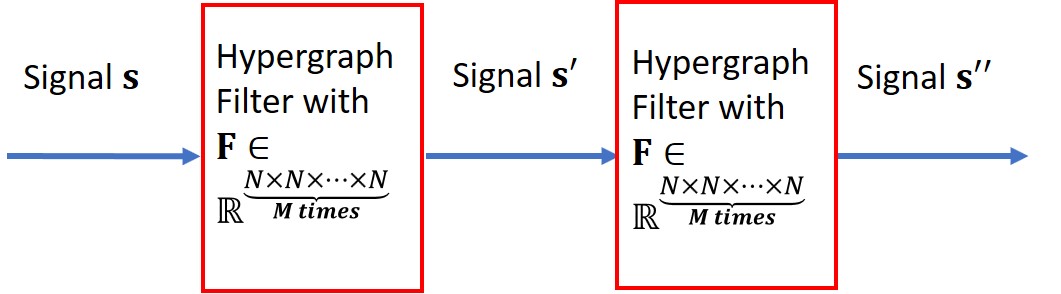}
	\caption{Signals in a Polynomial Filter.}
	\label{fil}
\end{figure}
In HGSP, we also relate signal element to one node in the hypergraph. Naturally, 
we can define the original signal as an $N$-length vector if there are $N$ 
nodes. Similarly as in GSP, we 
define the hypergraph shifting based on the representing tensor $\mathbf{F}$. 
However, since tensor $\mathbf{F}$ is of $M$-th order, 
we need an $(M-1)$-th order signal tensor to work with the 
hypergraph filter $\mathbf{F}$, such that the filtered signal is also 
an $N$-length vector as the original signal. 
For example, for a two-step polynomial filter shown as Fig. \ref{fil}, the signals $\mathbf{s,s',s''}$ should all be in the same dimension and order. 
For the input and output signals in a HGSP system to have a consistent form, we define an alternative form of the hypergraph signal as below.

\begin{myDef}[Hypergraph signal]
	For a hypergraph $\mathcal{H}$ with $N$ nodes and $m.c.e(\mathcal{H})=M$, 
an alternative form of hypergraph signal is an $(M-1)$-th order $N$-dimension tensor $\mathbf{s}^{[M-1]}$ obtained from $(M-1)$ times outer product of the original signal $\mathbf{s}=[s_1\quad s_2\quad...\quad s_N]^{\mathrm{T}}$, i.e.,
	\begin{equation}
	\mathbf{s}^{[M-1]}=\underbrace{\mathbf{s\circ...\circ s}}_{\text{M-1 times}},
	\end{equation}
	where each entry in position $(i_1,i_2, \cdots, i_{M-1})$ equals
the product $s_{i_1}s_{i_2}\cdots s_{i_{M-1}}$.
\end{myDef}

Note that the above hypergraph signal comes from the original signal. They are different forms of the same signal, which reflect the signal properties in different dimensions. For example, a second-order hypergraph signal highlights the properties of the two-dimensional 
signal components $s_is_j$ while the original signal directly emphasizes more about the one-dimension properties. We will discuss in greater details on 
the relationship between the hypergraph signal and the original 
signal in Section \uppercase\expandafter{\romannumeral3}-D. 

With the definition of hypergraph signals, let us define the original domain of signals for convenience before we step into the signal shifting. Similarly as that the signals lie in the time domain for DSP, we have the following definition of hypergraph vertex domain.

\begin{myDef}[Hypergraph vertex domain]
A signal lies in the hypergraph vertex domain if it resides on the structure of a hypergraph in the HGSP framework.
\end{myDef}
The hypergraph vertex domain is a counterpart of time domain in HGSP. The signals are analyzed based on the structure among vertices in a hypergraph.
\begin{figure}[t]
	\centering
	\includegraphics[width=3in]{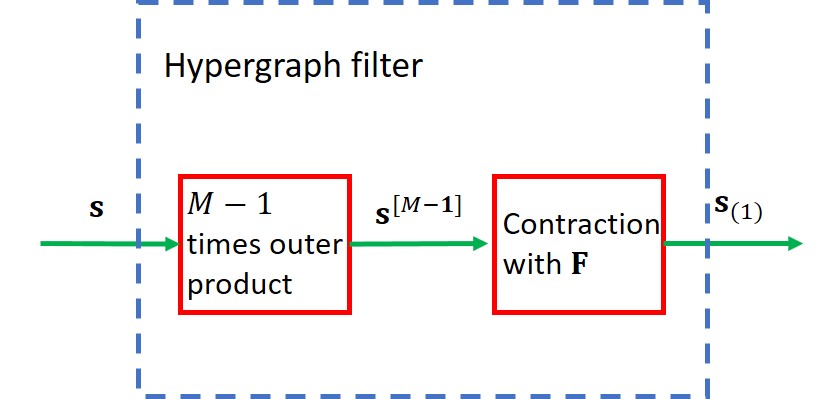}
	\caption{Diagram of Hypergraph Shifting.}
	\label{frame1}
\end{figure}

Next, we discuss how the signals shift on the given hypergraph. 
Recall that, in GSP, the signal shifting is defined by the product of the representing matrix $\mathbf{F_M}\in\mathbb{R}^{N\times N}$ and the signal vector $\mathbf{s}\in\mathbb{R}^{N}$, i.e., $\mathbf{s'=F_Ms}$. Similarly, we define the hypergraph signal shifting based on 
its tensor $\mathbf{F}$ and the hypergraph signal $\mathbf{s}^{[M-1]}$.

\begin{myDef}[Hypergraph shifting]
	The basic shifting filter of hypergraph signals is defined as the direct contraction between the representing tensor $\mathbf{F}$ and the hypergraph signals $\mathbf{s}^{[M-1]}$, i.e.,
	\begin{align}\label{11}
	\mathbf{s}_{(1)}=&\mathbf{F}\mathbf{s}^{[M-1]},
	\end{align}
where each element of the filter output is given by
	\begin{align}\label{com}
	(\mathbf{s}_{(1)})_i=\sum_{j_1,...,j_{M-1}=1}^{N} f_{ij_1...j_{M-1}}s_{j_1}s_{j_2}...s_{j_{M-1}}.
	\end{align}
\end{myDef}

Since the hypergraph signal contracts with the representing tensor in $M-1$ order, 
the one-time filtered signal $\mathbf{s}_{(1)}$ is an $N$-length vector, which has the same dimension as the original signal. Thus, the block diagram of a hypergraph filter 
with $\mathbf{F}$ can be shown as Fig. \ref{frame1}.

Let us now consider the functionality of the hypergraph filter, as well as the 
physical insight of the hypergraph shifting. 
In GSP, the functionality of the filter $\mathbf{F_M}$ is simply to shift
the signals along the link directions. However, interactions inside 
the hyperedge are more complex as it involves more than two nodes. 
In Eq. (\ref{com}), we see that the filtered signal in $\mathbf{v}_i$ equals 
the summation of the shifted signal components in all hyperedges containing node $\mathbf{v}_i$, 
where $f_{ij_1\cdots j_{M-1}}$ is the weight for each involved hyperedge and $\{s_{j_1},\cdots, s_{j_{M-1}}\}$ are the signals in the generalized hyperedges excluding $s_i$. 
Clearly, the hypergraph shifting multiplies signals in the same 
hyperedge of node $\mathbf{v}_i$ together before delivering the shift to 
a certain node $\mathbf{v}_i$. Taking the hypergraph in Fig. \ref{hyper3} 
as an example, node $\mathbf{v}_7$ is included in two hyperedges, $\mathbf{e}_2=\{\mathbf{v}_2,\mathbf{v}_3,\mathbf{v}_7\}$ and $\mathbf{e}_3=\{\mathbf{v}_5,\mathbf{v}_6,\mathbf{v}_7\}$. 
According to Eq. (\ref{com}), the shifted signal in node $\mathbf{v}_7$ 
is calculated as
\begin{equation}\label{22}
	s_7=f_{732}\times s_2s_3+f_{723}\times s_2s_3+f_{756}\times s_5s_6+f_{765}\times s_5s_6,
\end{equation}
where $f_{732}=f_{723}$ is the weight of the hyperedge $\mathbf{e}_2$ and $f_{756}=f_{765}$ is the weight for the hyperedge $\mathbf{e}_3$ in the adjacency tensor $\mathbf{F}$. 

\begin{figure}[t]
	\centering
	\subfigure[Example of signal shifting to node $\mathbf{v}_7$. Different colors of arrows show the different directions of shifting; `$\times$' refers to multiplication and `$+$' refers to summation.]{
		\label{ss3}
		\includegraphics[width=5cm]{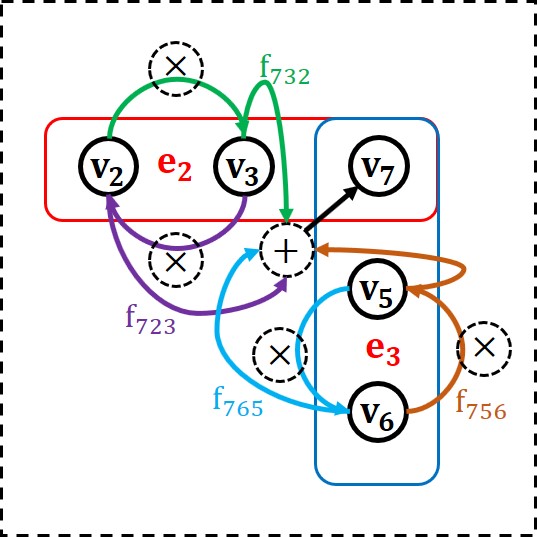}}
	\hspace{0.5in}
	\subfigure[Diagram of signal shifting to node $\mathbf{v}_1$ in an $M$-way hyperedge. Different colors refer to different shifting directions.]{
		\label{ss}
		\includegraphics[width=5cm]{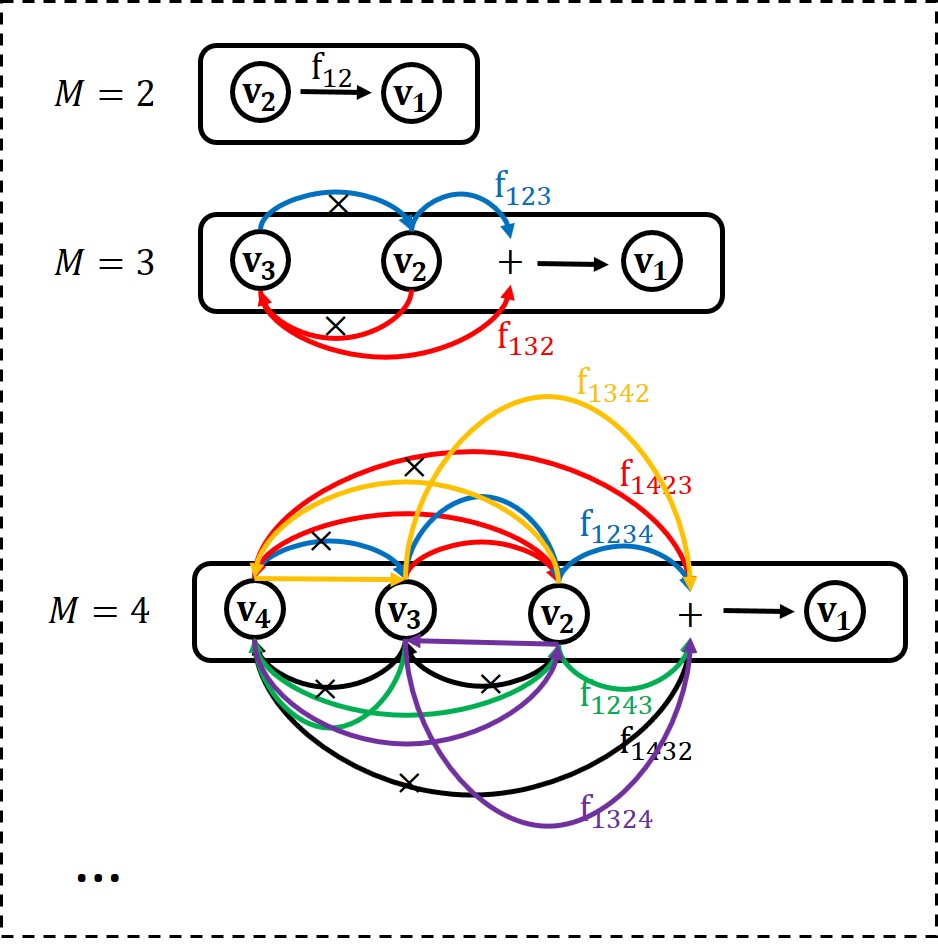}}
	\caption{Diagram of Signal Shifting.}
	\label{ss4}
\end{figure}

As the entry $a_{ji}$ in the
adjacency matrix of a normal graph indicates the link direction from the node $\mathbf{v}_i$ to the node $\mathbf{v}_j$, the entry $f_{i_1\cdots i_M}$ in the adjacency tensor 
similarly indicates the order of nodes in a hyperedge 
as $\{\mathbf{v}_{i_M},\mathbf{v}_{i_{M-1}},\cdots \mathbf{v}_{i_1}\}$, where $\mathbf{v}_{i_1}$ is the destination and $\mathbf{v}_{i_M}$ is the source. 
Thus, the shifting by Eq. (\ref{22}) could be interpreted as shown in
Fig. \ref{ss3}. Since 
there are two possible directions from nodes $\{\mathbf{v}_2,\mathbf{v}_3\}$ to 
node $\mathbf{v}_7$ in $\mathbf{e}_2$, 
there are two components shifted to $\mathbf{v}_7$, i.e., the 
first two terms in Eq. (\ref{22}). 
Similarly, there are also two components shifted by the hyperedge $\mathbf{e}_3$, 
i.e., the last two terms in Eq. (\ref{22}). 
To illustrate the hypergraph shifting more explicitly, Fig. \ref{ss} shows
a diagram of signal shifting to a certain node in an $M$-way hyperedge. 
From Fig. \ref{ss}, we see that the graph shifting in GSP is a special case of the
hypergraph shifting, where $M=2$. Moreover, 
there are $K=(M-1)!$ possible directions for the shifting to 
one specific node in an $M$-way hyperedge.

\subsection{Hypergraph Spectrum Space}
We now provide the definitions of the hypergraph Fourier space, i.e., the hypergraph spectrum space. In GSP, the graph Fourier space is defined as the eigenspace of its
representing matrix \cite{c4}. Similarly, we define the Fourier space of HGSP based on the representing tensor $\mathbf{F}$ of a hypergraph, 
which characterizes the hypergraph structure and signal shifting. For an $M$-th order $N$-dimension tensor $\mathbf{F}$, we can apply the orthogonal-CP decomposition \cite{c17} 
to write
\begin{equation}
\mathbf{F}\approx\sum_{r=1}^{R}\lambda_r\cdot\mathbf{f}_r^{(1)}\circ...\circ \mathbf{f}_r^{(M)},
\end{equation}
with basis $\mathbf{f}_r^{(i)}\in\mathbb{R}^N$ for $1\leq i\leq M$ and $\lambda_r\geq 0$.
Since $\mathbf{F}$ is super-symmetric \cite{c16}, i.e., $\mathbf{f}_r=\mathbf{f}_r^{(1)}=\mathbf{f}_r^{(2)}=\cdots=\mathbf{f}_r^{(M)}$, we have
\begin{equation}\label{14}
\mathbf{F}\approx\sum_{r=1}^{R}\lambda_r\cdot\underbrace{\mathbf{f}_r\circ...\circ \mathbf{f}_r}_{\text{M times}}.
\end{equation}
Generally, we have the rank $R\leq N$ in a hypergraph. We will discuss how to construct the 
remaining $\mathbf{f}_i$, $R<i\leq N$, for the case of $R< N$ later in Section III-F.

Now, by plugging Eq. (\ref{14}) into Eq. (\ref{11}), the hypergraph shifting can be written with the $N$ basis $\mathbf{f}_i$'s as
\begin{subequations}
\begin{align}
\mathbf{s}_{(1)}&=\mathbf{F}\mathbf{s}^{[M-1]}\\
&=(\sum_{r=1}^{N}\lambda_r\cdot\underbrace{\mathbf{f}_r\circ...\circ \mathbf{f}_r}_{\text{M times}})(\underbrace{\mathbf{s\circ...\circ s}}_{\text{M-1 times}})\\
&=\sum_{r=1}^{N}\lambda_r\mathbf{f}_r\underbrace{<\mathbf{f}_r,\mathbf{s}>\cdots<\mathbf{f}_r,\mathbf{s}>}_{\text{M-1 times}}\\
&=
\underbrace{
	\begin{bmatrix}
	\mathbf{f}_1& \cdots& \mathbf{f}_N
	\end{bmatrix}
	\begin{bmatrix}
	\lambda_1& & \\
	&\ddots& \\
	& &\lambda_N
	\end{bmatrix}
}_{\text{iHGFT and filter in Fourier space}}
\underbrace{
	\begin{bmatrix}
	(\mathbf{f}_1^{\mathrm{T}}\mathbf{s})^{M-1}\\
	\vdots\\
	(\mathbf{f}_N^{\mathrm{T}}\mathbf{s})^{M-1}
	\end{bmatrix}
}_{\text{HGFT of the hypergraph signal}},\label{20}
\end{align}
\end{subequations}
where $<\mathbf{f}_r,\mathbf{s}>=(\mathbf{f}_i^{\mathrm{T}}\mathbf{s})$ is 
the inner product between $\mathbf{f}_r$ and $\mathbf{s}$, and $(\cdot)^{M-1}$ is $(M-1)$th power.

From Eq. (\ref{20}), we see that the shifted signal in HGSP is in a similar decomposed to Eqs. (\ref{13}) and (\ref{44}) for GSP. The first two parts in Eq. (\ref{20}) work like $\mathbf{V_M^{-1}\Lambda}$ of the GSP eignen-decomposition, which could be interpreted as inverse Fourier transform and filter in the Fourier space. The third part can be understood as the hypergraph Fourier transform of the original signal. Hence, similarly as in GSP, 
we can define the hypergraph Fourier space and Fourier transform 
based on the orthogonal-CP decomposition of $\mathbf{F}$.

\begin{myDef}[Hypergraph Fourier space and Fourier transform]
	The hypergraph Fourier space of a given hypergraph $\mathcal{H}$ is defined as the space consisting of all orthogonal-CP decomposition basis $\{\mathbf{f}_1, \mathbf{f}_2,..., \mathbf{f}_N\}$. The frequencies are defined with respect
to the eigenvalue coefficients $\lambda_i$, $1\leq i \leq N$. 
The hypergraph Fourier transform (HGFT) of hypergraph signals is defined as
\begin{subequations}
	\begin{align}
	\mathbf{\hat s}&=\mathcal{F}_C(\mathbf{s}^{[M-1]})\\
	&=\begin{bmatrix}
	(\mathbf{f}_1^{\mathrm{T}}\mathbf{s})^{M-1}\\
	\vdots\\
	(\mathbf{f}_N^{\mathrm{T}}\mathbf{s})^{M-1}
	\end{bmatrix}.\label{33}
	\end{align}
\end{subequations}
\end{myDef}
Compared to GSP, if $M=2$, the HGFT has the same form as the traditional GFT. 
In addition, since $\mathbf{f}_r$ is the orthogonal basis, we have
\begin{equation}\label{eig}
\mathbf{Ff}_r^{[M-1]}=\sum \lambda_i \mathbf{f}_i(\mathbf{f}_i^{\mathrm{T}}\mathbf{f}_r)^{M-1}=\lambda_r \mathbf{f}_r.
\end{equation}
According to \cite{c16}, a vector $\mathbf{x}$ is an E-eigenvector of an $M$th-order tensor $\mathbf{A}$ if $\mathbf{Ax}^{[M-1]}=\lambda \mathbf{x}$ exists for a constant $\lambda$. 
Then, we obtain the following property of the hypergraph spectrum.
\begin{Pro}
	The hypergraph spectrum pair $(\lambda_r,\mathbf{f}_r)$ is an E-eigenpair of the representing tensor $\mathbf{F}$.
\end{Pro}

Recall that the spectrum space of GSP is the eigenspace of the representing matrix $\mathbf{F_M}$. \textit{Property 2} shows that HGSP has a consistent definition in the spectrum space as that for GSP.

\subsection{Relationship between Hypergraph Signal and Original Signal}
With HGFT defined, let us discuss more about the relationship between the hypergraph signal and the original signal in the Fourier space to understand the HGFT better. From Eq. (\ref{33}), the hypergraph signal in the Fourier space is written as 
\begin{align}
\mathbf{\hat s}=\begin{bmatrix}
(\mathbf{f}_1^{\mathrm{T}}\mathbf{s})^{M-1}\\
\vdots\\
(\mathbf{f}_N^{\mathrm{T}}\mathbf{s})^{M-1}
\end{bmatrix},
\end{align}
which can be further decomposed as
\begin{align}
\mathbf{\hat s}=\underbrace{(\begin{bmatrix}
	\mathbf{f}_1^{\mathrm{T}}\\
	\vdots\\
	\mathbf{f}_N^{\mathrm{T}}
	\end{bmatrix}\mathbf{s})*(\begin{bmatrix}
	\mathbf{f}_1^{\mathrm{T}}\\
	\vdots\\
	\mathbf{f}_N^{\mathrm{T}}
	\end{bmatrix}\mathbf{s})*\cdots*(\begin{bmatrix}
	\mathbf{f}_1^{\mathrm{T}}\\
	\vdots\\
	\mathbf{f}_N^{\mathrm{T}}
	\end{bmatrix}\mathbf{s})}_{M-1\quad times},\label{27}
\end{align}
where $*$ denotes Hadamard product.

From Eq. (\ref{27}), we see that the hypergraph signal in the hypergraph Fourier space is $M-1$ times Hadamard product of a component consisting 
of the hypergraph Fourier basis and the original signal. 
More specifically, this component works as the original signal in the hypergraph Fourier space, 
which is defined as 

\begin{equation}\label{sam2}
\mathbf{\tilde{s}}=\mathbf{Vs},	
\end{equation}
where $\mathbf{V}=[\mathbf{f}_1 \quad \mathbf{f}_2 \quad \cdots \quad \mathbf{f}_N]^{\mathrm{T}}$ and $\mathbf{V}^{\mathrm{T}}\mathbf{V=I}$.

Recall the definitions of the hypergraph signal and vertex domain in Section III-B, we have the following property.

\begin{Pro}
	The hypergraph signal is the $M-1$ times tensor outer product of the original signal in the hypergraph vertex domain, and the $M-1$ times Hadamard product of the original signal in the hypergraph frequency domain.
\end{Pro}

\begin{figure*}[t]
	\centering
	\subfigure[Process of HGFT]{
		\label{HGFT}
		\includegraphics[height=4cm,width=8cm]{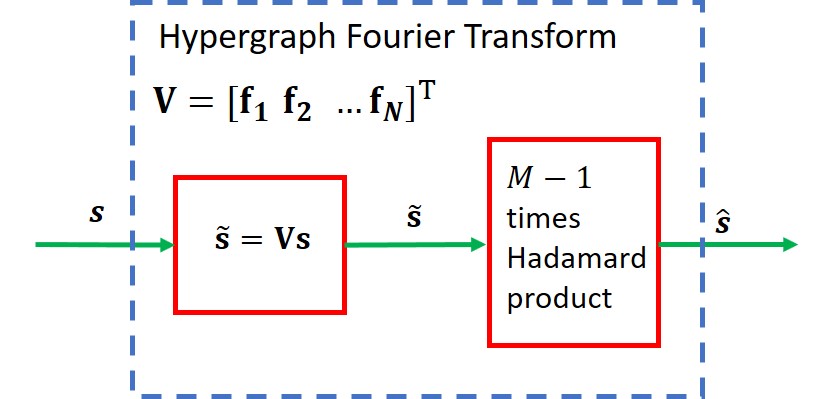}}
	\hspace{0.5in}
	\subfigure[Process of iHGFT]{
		\label{iHGFT}
		\includegraphics[height=4cm,width=8cm]{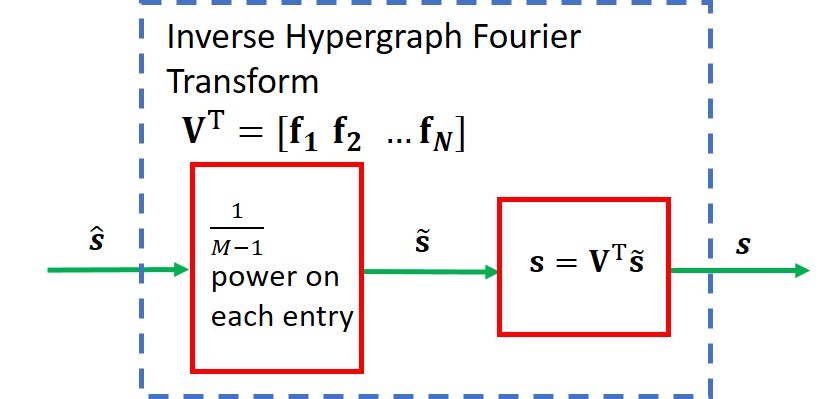}}
	\caption{Diagram of HGFT and iHGFT.}
	\label{relations1}
\end{figure*}

 Then, we could establish a connection between the original signal and the hypergraph signal in the hypergraph Fourier domain by the HGFT and inverse HGFT (iHGFT) as shown in Fig. \ref{relations1}. 
 Such a relationship 
 leads to some interesting properties and makes the HGFT implementation more straightforward, 
 which will be further discussed in Section III-F and Section III-G, respectively.

\subsection{Hypergraph Frequency}

As we now have a better understanding of the hypergraph Fourier space and Fourier transform,
we can discuss more about the hypergraph frequency and its order. 
In GSP, the graph frequency is defined with respect to the eigenvalues of the 
representing matrix $\mathbf{F_M}$ and ordered by the total variation \cite{c4}. 
Similarly, in HGSP, we define the frequency relative to the coefficients 
$\lambda_i$ from the orthogonal-CP decomposition. We order them by the total variation 
of frequency components $\mathbf{f}_i$ over the hypergraph.
The total variation of a general signal component over a hypergraph is defined as follows.
\begin{myDef}[Total variation over hypergraph]
	Given a hypergraph $\mathcal{H}$ with $N$ nodes and the normalized representing tensor $\mathbf{F}^{norm}=\frac{1}{\lambda_{max}}\mathbf{F}$, together with the original signal $\mathbf{s}$, the total variation over the hypergraph is defined as the total differences between the nodes and their corresponding neighbors in the perspective of shifting, i.e.,
\begin{subequations}
	\begin{align}
	\mathbf{TV}(\mathbf{\mathbf{s}})&=\sum_{i=1}^N|s_i-\sum_{j_1,\cdots ,j_{M-1}=1}^{N}F^{norm}_{ij_1\cdots j_{M-1}}s_{j_1}...s_{j_{M-1}}|\label{30}\\
	&=||\mathbf{s}-\mathbf{F}^{norm}\mathbf{s}^{[M-1]}||_1.
	\end{align}
\end{subequations}
\end{myDef}

We adopt the $l_1$-norm here only as an example of defining the total variation. 
Other norms may be more suitable depending on specific applications. 
Now, with the definition of total variation over hypergraphs, the frequency in HGSP is ordered by the total variation of the corresponding frequency component $\mathbf{f}_r$, 
i.e.,
\begin{equation}
\mathbf{TV(f}_r)=||\mathbf{f}_r-\mathbf{f}^{norm}_{r(1)}||_1,
\end{equation}
where $\mathbf{f}^{norm}_{r(1)}$ is the output of one-time shifting for $\mathbf{f}_r$ over the normalized representing tensor.

From Eq. (\ref{30}), we see that the total variation describes how much the signal component changes from a node to its neighbors over the hypergraph shifting. Thus, we have the following definition of hypergraph frequency.

\begin{myDef}[Hypergraph frequency]
Hypergraph frequency describes how oscillatory the signal component is with respect to the 
given hypergraph. A frequency component $\mathbf{f}_r$ is associated with a higher frequency if the total variation of this frequency component is larger.
\end{myDef}
Note that, the physical meaning of graph frequency was stated in GSP \cite{c1}. Generally, the graph frequency is highly related to the total variation of the corresponding frequency component. Similarly, the hypergraph frequency also relates to the corresponding total variation. We will discuss more about the interpretation of the hypergraph frequency and its relationships
with DSP and GSP later in Section IV-A, to further solidate our hypergraph frequency definition.

Based on the definition of total variation, we describe one important property of $\mathbf{TV(f}_r)$
in the following theorem.
\begin{theorem}
	\label{theoremA1}
	Define a supporting matrix
	\begin{equation}\label{sup}
	\mathbf{P_s}=\frac{1}{\lambda_{\max}}
	\begin{bmatrix}
	\mathbf{f}_1& \cdots& \mathbf{f}_N
	\end{bmatrix}
	\begin{bmatrix}
	\lambda_1& & \\
	&\ddots& \\
	& &\lambda_N
	\end{bmatrix}
	\begin{bmatrix}
	\mathbf{f}_1^{\mathrm{T}}\\
	\vdots\\
	\mathbf{f}_N^{\mathrm{T}}
	\end{bmatrix}.
	\end{equation}
With the normalized representing tensor $\mathbf{F}^{norm}=\frac{1}{\lambda_{\max}}\mathbf{F}$, the total variation of hypergraph spectrum $\mathbf{f}_r$ is calculated as
\begin{subequations}
	\begin{align}
	\mathbf{TV(f_r)}
	&=||\mathbf{f}_r-\mathbf{f}^{norm}_{r(1)}||_1\label{g1},\\
	&=||\mathbf{f}_r-\mathbf{P_s}\mathbf{f}_r||_1\label{g2},\\
	&=|1-\frac{\lambda_r}{\lambda_{\max}}|.\label{g3}
	\end{align}
\end{subequations}
	Moreover,  $\mathbf{TV}(\mathbf{f}_i)>\mathbf{TV}(\mathbf{f}_j)$ iff $\lambda_i<\lambda_j$.
\end{theorem}
\begin{IEEEproof}
	For hypergraph signals, the output of one-time shifting of $\mathbf{f}_r$ is calculated as
	\begin{equation}
	\mathbf{f}_{r(1)}=\sum_{i=1}^N \lambda_i \mathbf{f}_i(\mathbf{f}_i^{\mathrm{T}}\mathbf{f}_r)^{M-1}=\lambda_r \mathbf{f}_r. 
	\end{equation}
Based on the normalized $\mathbf{F}^{norm}$, we have $\mathbf{f}_{r(1)}^{norm}=\frac{\lambda_r}{\lambda_{max}}\mathbf{f}_r$. 
It is therefore 
easy to obtain Eq. (\ref{g3}) from Eq. (\ref{g1}).
To obtain Eq. (\ref{g2}), we have
	\begin{equation}
	\mathbf{P_sf}_r=\sum_{i=1}^N \lambda_i \mathbf{f}_i(\mathbf{f}_i^{\mathrm{T}}\mathbf{f}_r)=\frac{\lambda_r}{\lambda_{max}} \mathbf{f}_r. 
	\end{equation}
It is clear that Eq. (\ref{g2}) is the same as Eq. (\ref{g3}).
	
	Since $\lambda$ is real and nonnegative, we have
	\begin{equation}
		\mathbf{TV(f}_i)-\mathbf{TV(f}_j)=\frac{\lambda_j-\lambda_i}{\lambda_{max}}.
	\end{equation}
	Obviously, $\mathbf{TV}(\mathbf{f}_i)>\mathbf{TV}(\mathbf{f}_j)$ iff $\lambda_i<\lambda_j$.
\end{IEEEproof}

\textit{Theorem~\ref{theoremA1}} shows that the supporting matrix $\mathbf{P_s}$ can 
help us apply the total variation more efficiently in some real 
applications. Moreover, it provides the order of frequency
according to the coefficients $\lambda_i$'s with the following property.
\begin{Pro}
	A smaller $\lambda$ is related to a higher frequency in the hypergraph Fourier space, where its corresponding spectrum basis is called a high frequency component.
\end{Pro}

\subsection{Signals with Limited Spectrum Support}
With the order of frequency, we define the bandlimited signals as follows.

\begin{myDef}[Bandlimited signal]
	Order the coefficients as $\lambda=[\lambda_1\quad\cdots\quad\lambda_N]$ where $\lambda_1\geq\cdots\geq\lambda_{N}\geq0$, together with their corresponding $\mathbf{f}_r$'s. A hypergraph signal $\mathbf{s}^{[M-1]}$ is defined as $K$-bandlimited if the HGFT transformed signal $\mathbf{\hat{s}}=[\hat s_1, \cdots, \hat s_N]^{\mathrm T}$ has $\hat{s}_i=0$ for all $i\geq K$ where $K\in\{1,2,\cdots,N\}$. The smallest $K$ is defined as the bandwidth and the corresponding boundary is defined as $W=\lambda_K$. 
\end{myDef}

Note that, a larger $\lambda_i$ corresponds to a lower frequency as we mentioned in \textit{Property 4}. Then, the frequency are ordered from low to high in the definition above.
Moreover, we use the index $K$ instead of the coefficient 
value $\lambda$ to define the bandwidth for the following reasons:
\begin{itemize}
	\item Identical $\lambda$'s in two diferent hypergraphs do not refer to the same frequency. 
Since each hypergraph has its own adjacency tensor and spectrum space, the comparison of multiple spectrum pairs $(\lambda_i,\mathbf{f}_i)$'s is only meaningful within the same hypergraph. 
Moreover, there exists a normalization issue in the decomposition of different adjacency tensors. Thus, it is not meaningful to compare $\lambda_k$'s across 
two different hypergraphs.
	\item Since $\lambda_k$ values are not continuous over $k$,  different frequency cutoffs of $\lambda$ may lead to the same bandlimited space. For example, suppose that $\lambda_k=0.5$ and $\lambda_{k+1}=0.8$. Then, $\lambda=0.6$ and $\lambda'=0.7$ would lead to the same cutoff in
	the frequency space, which makes bandwidth definition non-unique.
\end{itemize}

As we discussed in Section III-D, the hypergraph signal is the Hadamard product of the original signal in the frequency domain. Then, we have the following property of bandwidth.
\begin{Pro}
	The bandwidth $K$ is the same based on the HGFT of the hypergraph signals $\mathbf{\hat s}$ and that of the original signals $\mathbf{\tilde s}$.
\end{Pro}
This property allows us to analyze the spectrum support of the hypergraph signal by looking into the original signal with lower complexity.
Recall that we can add $\mathbf{f}_i$ by using zero coefficients $\lambda_i$ 
when $R<N$ as mentioned in Section III-C. The added basis should not 
affect the HGFT signals in Fourier space. According to the structure of bandlimited signal, 
we need the added $\mathbf{f}_i$ could meet the following conditions: 
(1) $\mathbf{f}_i\perp \mathbf{f}_p$ for $p\neq i$; (2) $\mathbf{f}_i^\mathrm{T}\cdot\mathbf{s}\to 0$; and (3) $|\mathbf{f}_i|=1$.

\subsection{Implementation and Complexity}
We now consider the implementation and complexity issues of HGFT. 
Similar to GFT, the process of HGFT consists of two steps:
decomposition and execution. The decomposition is to 
calculate the hypergraph spectrum basis, and the execution transforms 
signals from the hypergraph vertex domain into the spectrum domain.

\begin{itemize}
	\item The calculation of spectrum basis by the orthogonal-CP decomposition is an important preparation step for HGFT. A straightforward algorithm would decompose the representing tensor $\mathbf{F}$ with the spectrum basis $\mathbf{f}_i$'s and coefficients $\lambda_i$'s as in
Eq. (\ref{14}). Efficient tensor decomposition is
an active topic in both fields of mathematics and engineering. 
There are a number of methods for CP decomposition in the literature.
In \cite{tds,td4}, motivated by the spectral theorem for real symmetric matrices, orthogonal-CP decomposition algorithms for symmetric tensors are developed based polynomial equations. In
\cite{c17},  Afshar \textit{et al.} proposed a more general decomposition algorithm for spatio-temporal data. Other works, including \cite{td1,td2,td3}, 
tried to develop faster decomposition methods 
for signal processing and big data applications. 
The rapid development of tensor decomposition and the advancement of computation ability 
will benefit the efficient derivation of hypergraph spectrum.

\item The execution of HGFT with a known spectrum basis is defined 
in Eq. (\ref{33}). 
According to Eq. (\ref{27}), the HGFT of hypergraph signal is an $M-1$ 
times Hadamard product of the original signal in the hypergraph spectrum space. 
This relationship can help execute HGFT and iHGFT of hypergraph signals 
more efficiently by applying matrix operations on the original signals. 
Clearly, the complexity of calculating the original signals in the frequency domain $\mathbf{\tilde s=Vs}$ is $O(N^2)$. 
In addition, since the computation complexity of the power function 
$x^{(M-1)}$ could be $O(\log (M-1))$ and each vector has $N$ entries, 
the complexity of calculating the $M-1$ times Hadamard product is $O(N\log(M-1))$. 
Thus, the complexity of general HGFT implementation is $O(N^2+N\log(M-1))$.
\end{itemize}

\section{Discussions and Interpretations}
In this section, we focus on the insights and physical meaning of frequency 
to help interpret the hypergraph spectrum space.
We also consider the relationships between HGSP and other existing 
works to better understand the HGSP framework.

\subsection{Interpretation of Hypergraph Spectrum Space}
We are interested in an intuitive interpretation of the hypergraph frequency 
and its relations with the DSP and GSP frequencies. 
We start with the frequency and the total variation in DSP. 
In DSP, the discrete Fourier transform (DFT) of a sequence $s_n$
is given by $\hat s_k=\sum_{n=0}^{N-1}s_n e^{-j\frac{2\pi kn}{N}}$ 
and the frequency is defined as $\nu_n=\frac{n}{N}$, $n=0,1, \cdots, N-1$. 
From \cite{dft}, we can easily summarize the following conclusions:
\begin{itemize}
	\item $\nu_n:\; 1<n<\frac{N}{2}-1$ corresponds to a continuous time
	signal frequency $ {n\over N} f_s$;
		\item $\nu_n:\; \frac{N}2+ 1< n<N-1$ corresponds to a continuous time
	signal frequency $- (1-{n\over N}) f_s$;
	\item $\nu_\frac{N}{2}$ corresponds to $f_s/2$; 
	\item $n=0$ corresponds to frequency 0. 
\end{itemize}
Here, $f_s$ is the critical sampling frequency.
In traditional DFT, we generate the Fourier transform $\hat f(\omega)=\int_{-\infty}^{\infty}f(x)e^{-2\pi jx\omega}dx$ at each discrete frequency $\frac{n}{N}f_s$, $n= -\frac{N}{2}+1, -\frac{N}{2}+2, \cdots, \frac{N}{2}-1, \frac{N}{2}$. The highest and lowest frequencies correspond to $n=N/2$ and $n=0$, respectively. Note that $n$ varies from $-\frac{N}{2}+1$ to $\frac{N}{2}$ here. Since $e^{-j2\pi k\frac{n}{N}}=e^{-j2\pi k\frac{n+N}{N}}$, we can let $n$ vary from $0$ to $N-1$ and cover the complete period. Now, $n$ varies in exact correspondence to $\nu_n$, and the aforementioned conclusions are drawn. The highest frequency occurs at $n=\frac{N}{2}$.

The total variation in DSP is defined as the differences among the signals over time \cite{tv}, i.e.,
\begin{subequations}
\begin{align}
\mathbf{TV(s)}&=\sum_{n=0}^{N-1}|s_{n}-s_{(n-1\mod N)}|\\&=\mathbf{||s-C_Ns||_1},
\end{align}
\end{subequations}
where
\begin{align}
{
\mathbf{C_N}=
\begin{bmatrix}
0&0&\cdots&0&1\\
1&0&\cdots&0&0\\
\vdots&\ddots&\ddots&\ddots&\vdots\\
0&0&\ddots&0&0\\
0&0&\cdots&1&0
\end{bmatrix}.}
\end{align}
 When 
we perform the eigen-decomposition of $\mathbf{C_N}$, we see that the 
eigenvalues are $\lambda_n=e^{-j\frac{2\pi n}{N}}$ with 
eigenvector $\mathbf{f}_n$, $0\leq n\leq N-1$.
More specifically, the total variation of the frequency component $\mathbf{f}_n$ is calculated as 
\begin{equation}
\mathbf{TV(f}_n)=|1-e^{j\frac{2\pi n}{N}}|,
\end{equation}
which increases with $n$ for $n\leq\frac{N}{2}$ before decreasing with $n$
for $\frac{N}{2}<n\le N-1$.

Obviously, the total variations of frequency components have a one-to-one 
correspondence to frequencies in the order of their values. 
If the total variation of a frequency component is larger, 
the corresponding frequency with the same index $n$ is higher. It 
also has clear physical meaning, i.e., 
a higher frequency component changes faster over time, which implies a larger total variation. 
Thus, we could also use the total variation of a frequency component to characterize its frequency 
in DSP.

\begin{figure}[t]
	\centering
	\includegraphics[height=2in]{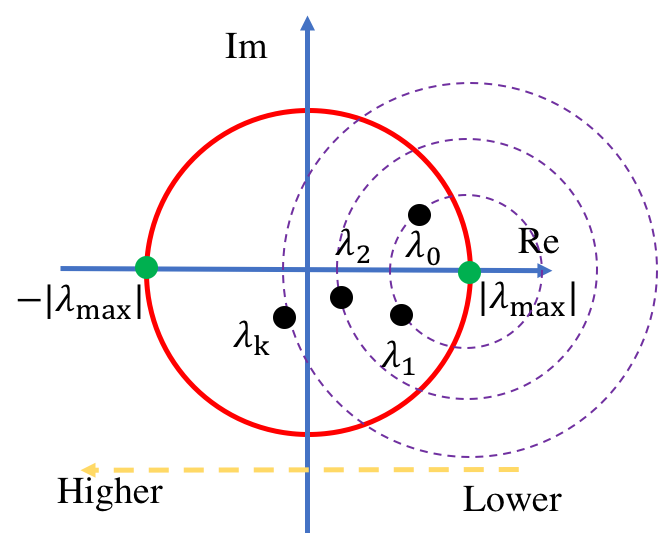}
	\caption{Example of Frequency Order in GSP for Complex Eigenvalues.}
	\label{fren}
\end{figure}

Let us now consider the total variation and frequency in GSP, where 
the signals are analyzed in the graph vertex domain instead of the time domain. 
Similar to the fact that the frequency in DSP describes the rate of signal changes over 
time, the frequency in GSP illustrates the rate of signal changes over vertex \cite{c4}. 
Likewise, the total variation of the graph Fourier basis defined according to 
the adjacency matrix $\mathbf{F_M}$ could be used to characterize each frequency. 
Since GSP handles signals in the graph vertex domain, the total variation of GSP is defined as the differences between all the nodes and their neighbors, i.e., 
\begin{subequations}
\begin{align}
\mathbf{TV}(\mathbf{\mathbf{s}})
&=\sum_{n=1}^N|s_n-\sum_{m}{F_M}^{norm}_{nm}s_m|\\
&=||\mathbf{s}-\mathbf{F_M}^{norm}\mathbf{s}||_1,
\end{align}
\end{subequations}
where $\mathbf{F_M}^{norm}=\frac{1}{|\lambda_{max}|}\mathbf{F_M}$.
If the total variation of the frequency component $\mathbf{{f_M}}_i$ is larger, it means the change over the graph between neighborhood vertices is faster, which indicates a higher graph frequency. Note that, once the graph is undirected, i.e., the eigenvalues are real numbers, the frequency decreases with the increase of the eigenvalue similar as HGSP in Section III-E; otherwise, if the graph is undirected, i.e., the eigenvalues are complex, the frequency changes as shown in Fig. \ref{fren}, which is consistency with the changing pattern of DSP frequency \cite{c4}.

\begin{figure}[t]
	\centering
	\includegraphics[height=2.3in]{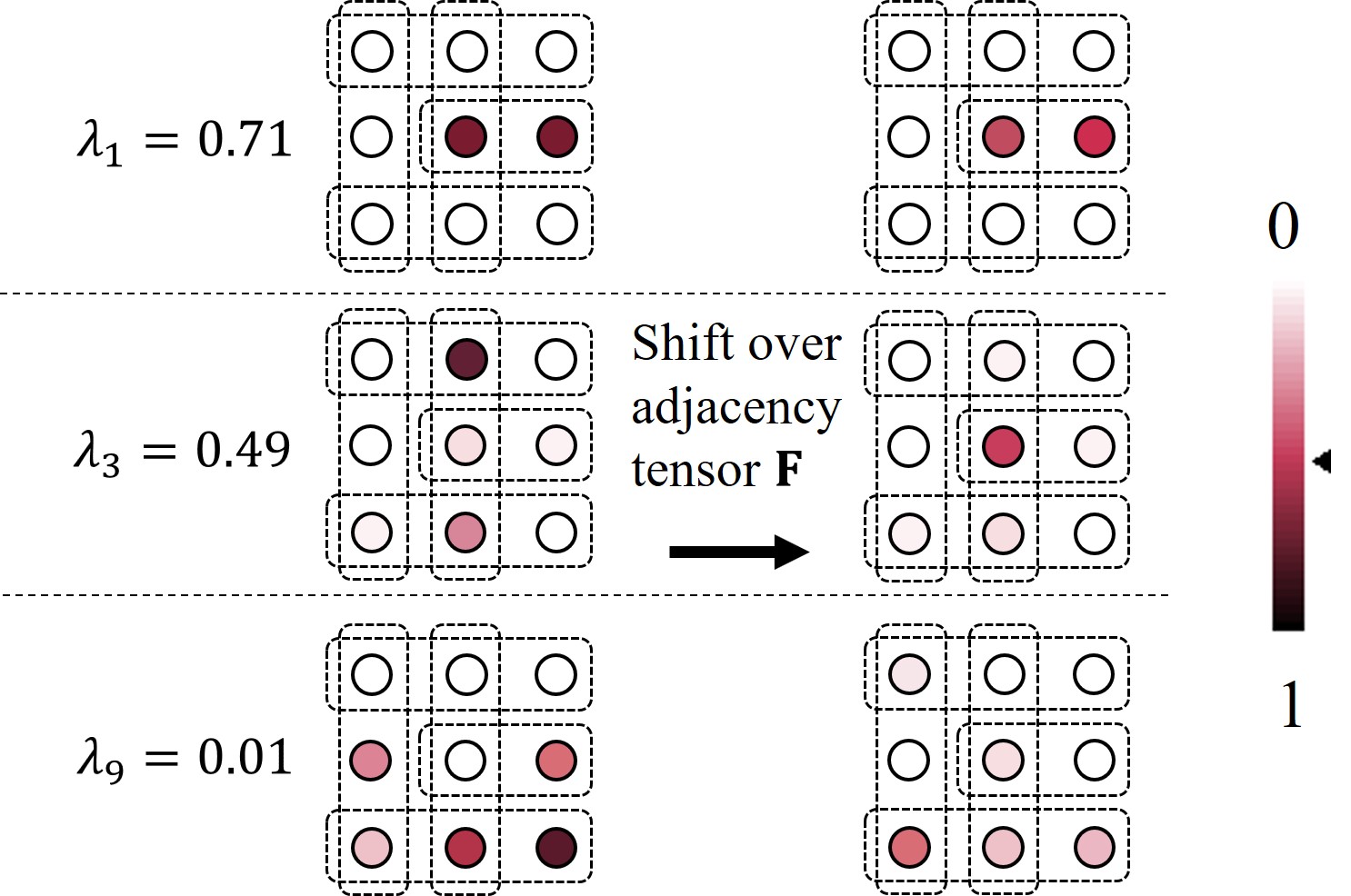}
	\caption{Frequency components in a hypergraph with 9 nodes and 5 hyperedges. The left panel shows the frequency components before shifting, and the right panel shows the frequency components after shifting. The values of signals are illustrated by colors. Higher frequency components imply larger changes across two panels.}
	\label{SSH1}
\end{figure}

We now turn to our HGSP framework. Like GSP, HGSP analyzes signals 
in the hypergraph vertex domain. Different from normal graphs, 
each hyperege in HGSP connects more than two nodes. The neighbors of 
a vertex $\mathbf{v}_i$ include all the nodes in the hyperedges containing $\mathbf{v}_i$. For example, if there exists a hyperedge $\mathbf{e}_1=\{\mathbf{v}_1,\mathbf{v}_2,\mathbf{v}_3\}$, nodes $\mathbf{v}_2$ and $\mathbf{v}_3$ are both neighbors of node $\mathbf{v}_1$. As we mentioned in Section III-E, the total variation of HGSP is defined as the difference between continuous signals over the hypergraph, i.e., the difference between the signal components and 
their respective shifted versions:
\begin{subequations}
\begin{align}
\mathbf{TV}(\mathbf{\mathbf{s}})&=\sum_{i=1}^N|s_i-\sum_{j_1,...,j_{M-1}}F^{norm}_{ij_1\cdots j_{M-1}}s_{j_1}\cdots s_{j_{M-1}}|\\
&=||\mathbf{s}-\mathbf{F}^{norm}\mathbf{s}^{[M-1]}||_1,
\end{align}
\end{subequations}
where $\mathbf{F}^{norm}=\frac{1}{\lambda_{max}}\mathbf{F}$.
Similar to DSP and GSP, pairs of $(\lambda_i, \mathbf{f}_i)$ in Eq. (\ref{14}) 
characterize the hypergraph spectrum space. A spectrum component with a
larger total variation represents a higher frequency component, 
which indicates faster changes over the hypergraph. Note that, as we mentioned in Section III-E, the total variation is larger and the frequency is higher if the corresponding $\lambda$ is smaller because we usually talk about undirected hypergraph and the $\lambda$'s are real in the tensor decomposition.
To illustrate it more clearly, we consider
a hypergraph with $9$ nodes, $5$ hyperedges, and $m.c.e=3$ as an example, 
shown in Fig. \ref{SSH1}. As we mentioned before, a smaller $\lambda$ indicates a higher frequency in HGSP. Hence, we see that the signals have more changes on each vertex if the frequency is higher.

\subsection{Connections to other Existing Works}
We now discuss the relationships between the HGSP and other existing works.

\subsubsection{Graph Signal Processing} 
One of the motivations for developing HGSP is to develop a more 
general framework for signal processing in high-dimensional graphs. 
Thus, GSP should be
a special case of HGSP. We illustrate the GSP-HGSP relationship as follows.
\begin{itemize}
	\item Graphical models: GSP is based on normal graphs\cite{c1}, where each simple 
edge connects exactly two nodes; HGSP is based on hypergraphs, where each hyperedge could connect more than two nodes. Clearly, the normal graph is a special case of hypergraph, 
for which the $m.c.e$ equals two. More specifically, a normal graph is a $2$-uniform hypergraph \cite{kunf}. Hypergraph provides a
more general model for multi-lateral relationships while normal graphs are only able to model bilateral relationship. 
For example, a $3$-uniform hypergraph is able to model the trilateral
interaction among users in a social network\cite{stag}. 
As hypergraph is a more general model for high-dimensional interactions, 
HGSP is also more powerful for high-dimensional signals.

\item Algebraic models: HGSP relies on tensors while GSP relies
on matrices, which are second-order tensors. 
Benefiting from the generality of tensor, HGSP is broadly 
applicable in high-dimensional data analysis.

\item Signals and signal shifting: In HGSP, we define the hypergraph signal as $M-1$ times tensor outer product of the original signal. 
More specifically, the hypergraph signal is the original signal if $M=2$. 
Basically, the hypergraph signal is the same as the graph signal 
if each hyperedge has exactly two nodes. 
Also shown in Fig. \ref{ss} of Section III-C, 
graph shifting is a special case of hypergraph shifting when $M=2$.
	
\item Spectrum properties: In HGSP, the spectrum space is defined
over the orthogonal-CP decomposition in terms of the basis and coefficients, which are also the E-eigenpairs of the representing tensor \cite{teig}, shown in Eq. (\ref{eig}). 
In GSP, the spectrum space is defined as the matrix eigenspace. 
Since the tensor algebra is an extension of matrix, the
HGSP spectrum is also an extension of the GSP spectrum. 
For example, as discussed in Section III, GFT is the same as 
HGFT when $M=2$.
\end{itemize}
Overall, HGSP is an extension of GSP, which is both more general and novel.
The purpose of developing the HGSP framework is to 
facilitate more interesting signal processing tasks 
that involve high-dimensional signal interactions.

\subsubsection{Higher-Order Statistics}
Higher-order statistics (HOS) has been effectively applied 
in signal processing\cite{hos1,hos2}, which can analyze the multi-lateral interactions of signal samples 
and have found successes in many applications, such as blind feature detection 
\cite{hos3}, decision \cite{hos4}, and signal classifications \cite{hos5}. 
In HOS, the $k$th-order cumulant of random variables $\mathbf{x}=[x_1,\cdots,x_k]^{\mathrm{T}}$ is defined \cite{hos7} based on the coefficients of $\mathbf{v}=[v_1,\cdots,v_k]^{\mathrm{T}}$ in the Talyor series expansion of cumulant-gernerating function, i.e.,
\begin{equation}\label{cum1}
	K(\mathbf{v})=\ln \mathbf{E}\{\exp(j\mathbf{v}^{\mathrm{T}}\mathbf{x})\}.
\end{equation}

It is easy to see that HGSP and HOS are related in high-dimensional signal processing. 
They can be both represented by tensor. For example, 
in the multi-channel problems of \cite{hos8}, the 3rd-order
cumulant $\mathbf{C}=\{C_{y_i,y_j,y_z}(t,t_1,t_2)\}$ of zero-mean signals can be represented as a multilinear array, e.g.,
\begin{equation}
	 C_{y_i,y_j,y_z}(t,t_1,t_2)=\mathbf{E}\{y_i(t)y_j(t+t_1)y_z(t+t_2)\},
\end{equation}
which is essentially a third-order tensor. 
More specifically, if there are $k$ samples, the cumulant $\mathbf{C}$ 
can be represented as an $p^k$-element vector, 
which is the flattened signal tensor similar to 
the $n$-mode flattening of HGSP signals. 

Although both HOS and HGSP are high-dimensional signal processing tools, 
they focus on complementary aspects of the signals. Specifically, HGSP aims to analyze signals 
over the high-dimensional vertex domain, while HOS focuses on the statistical domain. 
In addition, the forms of signal combination are also different, where 
HGSP signals are based on the hypergraph shifting defined as in
Eq. (\ref{com}), whereas HOS cumulants 
are based on the statistical average of shifted signal products.

\subsubsection{Learning over Hypergraphs}
Hypergraph learning is another tool to handle 
structured data and sometimes uses similar techniques to HGSP. 
For example, the authors of \cite{hgl1} proposed an alternative definition of 
hypergraph total variation and design algorithms 
in accordance for classification and clustering problems. In addition,
hypergraph learning also has its own definition of
the hypergraph spectrum space. For example, \cite{c29,c30} 
represented the hypergraphs using a graph-like similarity matrix 
and defined a spectrum space as the eigenspace of this similarity matrix. 
Other works considered different aspects of hypergraph, 
including the hypergraph Laplacian \cite{hgl2} and hypergraph lifting \cite{hlift}. 

The HGSP framework exhibits features different from hypergraph learning: 
1) HGSP defines a framework 
that generalizes the classical digital signal processing and 
traditional graph signal processing; 2) HGSP applies different definitions of hypergraph
characteristics such as the total variation, spectrum space, and Laplacian; 3) HGSP 
cares more about the spectrum space while learning focuses more on data; 
4) As HGSP is an extension of DSP and GSP, 
it is more suitable to handle detailed tasks such as compression, denoising, 
and detection. All these features make HGSP a different technical concept 
from hypergraph learning.

\section{Tools for Hypergrph Signal Processing}
In this section, we introduce several useful tools built within the framework of HGSP.

\subsection{Sampling Theory}
Sampling is an important tool in data analysis, which selects a subset of individual
data points to estimate the characteristics of the whole population \cite{sam}. 
Sampling plays an important role in applications such as
compression \cite{C18} and storage \cite{sam1}. 
Similar to sampling signals in time, the HGSP sampling theory can be 
developed to sample signals over the vertex domain. 
We now introduce the basics of HGSP sampling theory for lossless 
signal dimension reduction. 

To reduce the size of a hypergraph signal $\mathbf{s}^{[M-1]}$, 
there are two main approaches: 1) to reduce the dimension of each order; and 2)
to reduce the number of orders. Since the 
reduction of order breaks the structure of hypergraph and cannot always guarantee perfect recovery, we adopt the dimension reduction of each order. 
To change the dimension of a certain order, we can use the $n$-Mode product. 
Since each order of the hypergraph signal is equivalent, the $n$-Mode product 
operators of each order are the same. 
Then, the sampling operation of the hypergraph signal is defined as follows:
\begin{myDef}[Sampling and Interpolation]
	Suppose that $Q$ is the dimension of each sampled order. The sampling operation is defined as
	\begin{equation}
	\mathbf{s_Q^{[M-1]}}=\mathbf{s^{[M-1]}}\times_1\mathbf{U}\times_2\mathbf{U}\cdots\times_{M-1}\mathbf{U},
	\end{equation}
	where the sampling operator is $\mathbf{U}\in\mathbb{R}^{Q\times N}$ to be defined later, and the sampled signal is $\mathbf{s_Q^{[M-1]}}\in \mathbb{R}^{\underbrace{\scriptstyle{Q\times Q\times...\times Q}}_{{M-1}\  \rm{ times}}}$.
	
	The interpolation operation is defined by 
	\begin{equation}
	\mathbf{s^{[M-1]}}=\mathbf{s_Q^{[M-1]}}\times_1\mathbf{T}\times_2\mathbf{T}
	\cdots\times_{M-1}\mathbf{T},
	\end{equation}
	where the interpolation operator is $\mathbf{T}\in\mathbb{R}^{N\times Q}$ to be defined later.
\end{myDef}

As presented in Section III, the hypergraph signal and original signal 
are different forms of the same data. They may have similar properties 
in structures. To derive the sampling theory for perfect signal recovery efficiently, 
we first consider the sampling operations of the original signal.

\begin{myDef}[Sampling original signal]
	Suppose an original $K$-bandlimited signal $\mathbf{s}\in \mathbb{R}^N$ is 
to be sampled into $\mathbf{s_Q}\in\mathbb{R}^Q$, where $q=\{q_1,\cdots ,q_Q\}$ denotes the sequence of sampled indices and $q_i\in\{1,2,\cdots,N\}$. The sampling operator $\mathbf{U}\in\mathbb{R}^{Q\times N}$ is a linearing mappling from $\mathbb{R}^N$ to $\mathbb{R}^Q$, defined by
	\begin{equation}
	U_{ij}=\left\{
	\begin{aligned}
	1,&\quad j=q_i; \\
	0,&\quad \mbox{otherwise}, \\
	\end{aligned}
	\right.
	\end{equation}
	and the interpolation operator $\mathbf{T}\in\mathbb{R}^{N\times Q}$ is a linear mapping from $\mathbb{R}^Q$ to $\mathbb{R}^N$. 
	Then, the sampling operation is defined by
	\begin{equation}
	\mathbf{s_Q=U\cdot s},
	\end{equation}
	and the interpolation operation is defined by
	\begin{equation}
	\mathbf{s'=T\cdot s_Q}.
	\end{equation}
\end{myDef}

Analyzing the structure of the sampling operations, we have the following properties.
\begin{theorem}
	The hypergraph signal $\mathbf{s^{[M-1]}}$ shares the same sampling operator $\mathbf{U}\in\mathbb{R}^{Q\times N}$ and interpolation operator $\mathbf{T}\in\mathbb{R}^{N\times Q}$ with the original signal $\mathbf{s}$.
\end{theorem}

\begin{IEEEproof} We first examine one of the orders in $n$-Mode product 
	of hypergraph signal, i.e., $n$th-order of $\mathbf{s^{[M-1]}}$, $1\leq n \leq N$, as
	\begin{equation}\label{sam1}
	(\mathbf{s^{[M-1]}}\times_n\mathbf{U})_{i_1...i_{n-1}ji_{n+1}...i_{M-1}}=\sum_{i_n=1}^{N}s_{i_1}s_{i_2}...s_{i_{M-1}}U_{ji_n}.
	\end{equation}
	Since all elements in $\mathbf{s_Q^{[M-1]}}$ should also be the elements of $\mathbf{s^{[M-1]}}$ after sampling, only one $U_{ji_n}=1$ exists for each $j$ according to Eq. (\ref{sam1}), i.e., only one term in the summation exists for each $j$ in the right part of Eq. (\ref{sam1}). Moreover, since $\mathbf{U}$ samples over all the order, $U_{pi_n}=1$ and $U_{ji_n}=1$ cannot exist at the same time so that all the entries in $\mathbf{s_Q^{[M-1]}}$ are also in $\mathbf{s^{[M-1]}}$. Suppose $q=\{q_1,q_2,\cdots,q_Q\}$ is the places of non-zero $U_{jq_j}$'s, we have
	\begin{equation}
	\mathbf{s_Q^{[M-1]}}(i_1,i_2,\cdots,i_Q)=s_{i_{q_1}}s_{i_{q_2}}\cdots s_{i_{q_Q}}.
	\end{equation}
As a result, we have
$
	U_{ji}=\delta[i-q_j]
$,
	which is the same as the sampling operator for the original signal. For the interpolation operator, the proof is similar and hence omitted. 
\end{IEEEproof}

Given \textit{Theorem 2}, we only need to analyze the operations of the original signal in the sampling theory. Next, we discuss the conditions for perfect recovery.
For the original signal, we have the following property.
\begin{lemma}
	Suppose that $\mathbf{s}\in\mathbb{R}^N$ is a $K$-bandlimited signal. Then, we have
	\begin{equation}
	\mathbf{s}=\mathcal{F}^\mathrm{T}_{[K]}\mathbf{\tilde{s}}_{[K]},
	\end{equation}
	where $\mathcal{F}^\mathrm{T}_{[K]}=[\mathbf{f}_1,\cdots,\mathbf{f}_K]$ and $\mathbf{\tilde{s}}_{[K]}\in \mathbb{R}^K$ consists of the first $K$ elements of the original signal in the frequency domain, i.e., $\mathbf{\tilde{s}}$.
\end{lemma}
\begin{IEEEproof}
	Since $\mathbf{s}$ is $K$-bandlimited, $\mathbf{\tilde s}_i=\mathbf{f}_i^\mathrm{T}\mathbf{s}=0$ when $i>K$.
	Then, according to Eq.(\ref{sam2}), we have
	\begin{align}
	\mathbf{s}=\mathbf{V}^\mathrm{T}\mathbf{Vs}
	=\sum_{i=1}^{K}\mathbf{f}_i\mathbf{f}_i^\mathrm{T}\mathbf{s}+\sum_{i=K+1}^{N}\mathbf{f}_i\mathbf{f}_i^\mathrm{T}\mathbf{s}
	=\sum_{i=1}^{K}\mathbf{f}_i\mathbf{f}_i^\mathrm{T}\mathbf{s}+0
	=\mathcal{F}^\mathrm{T}_{[K]}\mathbf{\tilde{s}}_{[K]},
	\end{align}
	where $\mathbf{V}=[\mathbf{f}_1,\cdots,\mathbf{f}_N]^\mathrm{T}$.
\end{IEEEproof}

This lemma implies that the first $K$ frequency components carry all the information of the original signal. Since the hypergraph signal and the original signal share the same sampling operators, we can reach a similar conclusion for perfect recovery as \cite{C18, C19}, given in the following theorem.
\begin{theorem}
	Define the sampling operator $\mathbf{U}\in\mathbb{R}^{Q\times N}$ according to
$
	U_{ji}=\delta[i-q_j]$
where $1\le q_i\le N, \; i=1,\ \ldots,\ Q$. By choosing $Q\geq K$ and the interpolation operator $\mathbf{T}=\mathcal{F}^\mathrm{T}_{[K]}\mathbf{Z}\in\mathbb{R}^{N\times Q}$ with $\mathbf{ZU}\mathcal{F}^\mathrm{T}_{[K]}=\mathbf{I_K}$ and $\mathcal{F}^\mathrm{T}_{[K]}=[\mathbf{f}_1,\cdots,\mathbf{f}_K]$, we can achieve a perfect recovery, i.e., $\mathbf{s=TUs}$ for all $K$-bandlimited original signal $\mathbf{s}$ and the corresponding hypergraph signal $\mathbf{s}^{[M-1]}$.
\end{theorem}

\begin{IEEEproof}
	To prove the theorem, we show that $\mathbf{TU}$ is a projection operator 
and $\mathbf{T}$ spans the space of the first $K$ eigenvectors. 
	From \textit{Lemma 1} and $\mathbf{s=Ts_Q}$, we have
	\begin{equation}
	\mathbf{s}=\mathcal{F}^\mathrm{T}_{[K]}\mathbf{\tilde{s}}_{[K]}=\mathcal{F}^\mathrm{T}_{[K]}\mathbf{Zs_Q}.
	\end{equation}
As a result, $
	rank(\mathbf{Zs_Q})=rank(\mathbf{\tilde{s}}_{[K]})=K.
$ Hence, we conclude that $K\leq Q$. 
	
Next, we show that $\mathbf{TU}$ is a projection by proving that
$	\mathbf{TU\cdot TU=TU}.$
Since we have $Q\geq K$ and
	\begin{equation}\label{sam3}
\mathbf{ZU}\mathcal{F}^\mathrm{T}_{[K]}=\mathbf{I_K},
	\end{equation}
We have 
\begin{subequations}
\begin{align}
\mathbf{TU\cdot TU}&=\mathcal{F}^\mathrm{T}_{[K]}\mathbf{ZU}\mathcal{F}^\mathrm{T}_{[K]}\mathbf{ZU}\\
&= \mathcal{F}^\mathrm{T}_{[K]}\mathbf{ZU} = \mathbf{TU}.
\end{align}	
\end{subequations}
Hence, TU is a projection operator. 
For the spanning part, the proof is the same as that in \cite{C18}.
\end{IEEEproof}

\textit{Theorem 3} shows that a perfect recovery is possible for a 
bandlimited hypergraph signal. We now examine some interesting properties of the
sampled signal.

From the previous discussion, we have $\mathbf{\tilde{s}}_{[K]}=\mathbf{Zs_Q}$, which has a 
similar form to HGFT, where $\mathbf{Z}$ can be treated as the Fourier transform operator. 
Suppose that $Q=K$ and $\mathbf{Z}=[\mathbf{z}_1\quad \cdots\quad \mathbf{z}_K]^\mathrm{T}$. We have the following first-order difference property.
\begin{theorem}
	Define a new hypergraph by $\mathbf{F_K}=\sum_{i=1}^{K}\lambda_i\cdot\mathbf{z}_i\circ\cdots\circ\mathbf{z}_i$. 
	Then, for all $K$-bandlimited signal $\mathbf{s}^{[M-1]}\in\mathbb{R}^{\underbrace{\scriptstyle{N\times N\times...\times N}}_{\text{M times}}}$, 
	it holds that
	\begin{equation}\label{sam4}
	\mathbf{s}_{[K]}-\mathbf{F_Ks}^{[M-1]}_{[K]}=\mathbf{U}(\mathbf{s-Fs}^{[M-1]}).
	\end{equation}
\end{theorem}

\begin{IEEEproof}
	Let the diagonal matrix $\Sigma_{[K]}$ consist of the
 first $K$ coefficients $\{\lambda_1,\;\ldots,\;\lambda_K\}$.
	Since $\mathbf{ZU}\mathcal{F}^\mathrm{T}_{[K]}=\mathbf{I_K}$, we have
	\begin{subequations}
	\begin{align}
	\mathbf{F_Ks}^{[M-1]}_{[K]}&=\mathbf{Z}^{-1}\Sigma_{[K]}\mathbf{\hat s}_{[K]}\\
	&=\mathbf{U}\mathcal{F}^\mathrm{T}_{[K]}\Sigma_{[K]}\mathbf{\hat s}_{[K]}\\
	&=\mathbf{U}[(\sum_{i=1}^{K}\lambda_i\cdot\underbrace{\mathbf{f}_i\circ...\circ \mathbf{f}_i}_{\text{M times}})(\underbrace{\mathbf{s\circ...\circ s}}_{\text{M-1 times}})+0]\\
	&=\mathbf{UFs}^{[M-1]}.
	\end{align}
\end{subequations}
Since $\mathbf{s}_{[K]}=\mathbf{Us}$, it therefore holds that $\mathbf{s}_{[K]}-\mathbf{F_Ks}^{[M-1]}_{[K]}=\mathbf{U}(\mathbf{s-Fs}^{[M-1]}).$
\end{IEEEproof}

\textit{Theorem 4} shows that the sampled signals form a new hypergraph that preserves 
the information of the one-time shifting filter over the original hypergraph. 
For example, the left-hand side
of Eq. (\ref{sam4}) represent the difference between the
sampled signal and the one-time shifted version in the new hypergraph. 
The right-hand side of Eq. (\ref{sam4}) is the difference between a signal and 
its one-time shifted version in the original hypergraph, together with the sampling operator. 
That is, the sampled result of the one-time shifting differences 
in the original hypergraph is equal to the one-time shifting differences 
in the new sampled hypergraph.

\subsection{Filter Desgin}
Filter is an important tool in signal processing applications such as
denoising, feature enhancement, smoothing, and classification. 
In GSP, the basic filtering is defined as $\mathbf{s'=F_Ms}$ where 
$\mathbf{F_M}$ is the representing matrix\cite{c1}. In HGSP, the basic hypergraph filtering 
is defined in Section III-C as $\mathbf{s}_{(1)}=\mathbf{Fs}^{[M-1]}$, which is designed
according to the tensor contraction. 
The HGSP filter is a multilinear mapping \cite{fl1}. 
The high-dimensionality of tensors provides
more flexibility in designing the HGSP filter.

\subsubsection{Polynomial Filter based on Representing Tensor}
Polynomial filter is one basic form of HGSP filters, with 
which signals are shifted several times over the hypergraph. 
An example of polynomial filter is given as Fig. \ref{fil} in Section III-B. 
A $k$-time shifting filter is defined as
\begin{subequations}
\begin{align}
	\mathbf{s}_{(k)}&=\mathbf{F}\mathbf{s}_{(k-1)}^{[M-1]}\\
	&=\underbrace{ \mathbf{F}(\mathbf{F}(...(\mathbf{Fs}^{[M-1]})^{[M-1]})^{[M-1]})^{[M-1]}}_{k\quad times}.
\end{align}
\end{subequations}
More generally, a polynomial filter is designed as
\begin{equation}\label{poly1}
	\mathbf{s'}=\sum_{k=1}^a \alpha_k \mathbf{s}_{(k)},
\end{equation}
where $\{\alpha_k\}$ are the filter coefficients. 
Such HGSP filters are based on multilinear tensor contraction, which could be used for different signal processing tasks by selecting
specific parameters $a$ and $\{\alpha_i\}$.

In addition to the general polynomial filter based on hypergraph signals, 
we provide another specific form of polynomial filter based on the original signals. 
As mentioned in Section III-E, the supporting matrix $\mathbf{P_s}$ in Eq. (\ref{sup}) 
captures all the information of the frequency space. For example, 
the unnormalized supporting matrix $\mathbf{P}=\lambda_{max}\mathbf{P_s}$ is calculated as
\begin{equation}\label{P}
\mathbf{P}=
\begin{bmatrix}
\mathbf{f}_1& \cdots& \mathbf{f}_N
\end{bmatrix}
\begin{bmatrix}
\lambda_1& & \\
&\ddots& \\
& &\lambda_N
\end{bmatrix}
\begin{bmatrix}
\mathbf{f}_1^{\mathrm{T}}\\
\vdots\\
\mathbf{f}_N^{\mathrm{T}}
\end{bmatrix}.
\end{equation}
Obviously, the hypergraph spectrum pair $(\lambda_r,\mathbf{f}_r)$ is an eigenpair of the supporting matrix $\mathbf{P}$. Moreover, \textit{Theorem 1} shows that the total variation of frequency component equals to a function of $\mathbf{P}$, i.e.,
\begin{equation}\label{tv1}
	\mathbf{TV(f}_r)=||\mathbf{f}_r-\frac{1}{\lambda_{max}}\mathbf{Pf}_r||_1. 
\end{equation}
From Eq. (\ref{tv1}), $\mathbf{P}$ can be interpreted as a shifting matrix for the original signal. 
Accordingly, we can design a polynomial filter for the original signal
based on the supporting matrix $\mathbf{P}$ whose
$k$th-order term is defined as 
\begin{equation}\label{poly2}
	\mathbf{s}_{<k>}=\mathbf{P}^k\mathbf{s}.
\end{equation}
The $a$-th order polynomial filter is simply given as
\begin{equation}
	\mathbf{s}'=\sum_{k=1}^{a}\alpha_k\mathbf{P}^k\mathbf{s}.
\end{equation}
A polynomial filter over the original signal can be 
determined with specific choices of $a$ and $\alpha$.

Let us consider some interesting properties of the polynomial filter for the original signal. 
First, given the $k$th-order term, we have the following property as \textit{Lemma 2}.
\begin{lemma}
	\begin{equation}
	\mathbf{s}_{<k>}=\sum_{r=1}^N \lambda_r^k  (\mathbf{f}_r^{\mathrm{T}}\mathbf{s})\mathbf{f}_r.
	\end{equation}
\end{lemma}

\begin{IEEEproof}
Let $\mathbf{V}^\mathrm{T}=[\mathbf{f}_1,\cdots,\mathbf{f}_N]$ and $\Sigma=diag([\lambda_1,\cdots,\lambda_N])$. Since
$\mathbf{V}^\mathrm{T}\mathbf{V}=\mathbf{I}$, 
we have
\begin{subequations}
\begin{align}
	\mathbf{P}^k&=\underbrace{\mathbf{V}^\mathrm{T}\Sigma\mathbf{V}\mathbf{V}^\mathrm{T}\Sigma\mathbf{V}\cdots \mathbf{V}^\mathrm{T}\Sigma\mathbf{V}}_{k\quad times}\\
	&=\mathbf{V}^\mathrm{T}\Sigma^k\mathbf{V}.
\end{align}\end{subequations}

Therefore, the $k$th-order term is given as
\begin{subequations}
\begin{align}
	\mathbf{s}_{<k>}&=\mathbf{V}^\mathrm{T}\Sigma^k\mathbf{V}\mathbf{s}\\
	&=\sum_{r=1}^N \lambda_r^k  (\mathbf{f}_r^{\mathrm{T}}\mathbf{s})\mathbf{f}_r.
\end{align}
\end{subequations}
\end{IEEEproof}

From \textit{Lemma 2}, we obtain the following property of the polynomial filter for the original signal.
\begin{theorem} Let $h(\cdot)$ be a polynomial function.
	For the polynomial filter $\mathbf{H}=h(\mathbf{P})$ for the original signal, the filtered signal 
	satisfies
\begin{align}
	\mathbf{Hs}=\sum_{r=1}^{N} h(\mathbf{s}_{<i>})
=\sum_{r=1}^{N} h(\lambda_r) \mathbf{f}_r (\mathbf{f}_r^{\mathrm{T}}\mathbf{s}).
	\end{align}
\end{theorem}
This theorem works as the invariance property of exponential in HGSP, similar to those in GSP and DSP \cite{c1}. Eq. (\ref{poly1}) and Eq. (\ref{poly2}) provide more choices for HGSP polynomial filters in hypergraph signal processing and data analysis. We will give specific examples of 
practical applications in Section VI.

\subsubsection{General Filter Design based on Optimization}
In GSP, some filters are designed via optimization formulations \cite{c1,fl2,fl3}. Similarly, general HGSP filters 
can also be designed via optimization approaches. 
Assume $\mathbf{y}$ is the oberserved signal before shifting and $\mathbf{s}=h(\mathbf{F,y})$ is the shifted signal by HGSP filter $h(\cdot)$ designed for specific applications.
Then, the filter design can be formulated as
\begin{equation}
	\min_{h} ||\mathbf{s-y}||^2_2+\gamma f(\mathbf{F,s}),
\end{equation}
where $\mathbf{F}$ is the representing tensor of the hypergraph and $f(\cdot)$ is a 
penalty function designed for specific problems. For example, 
the total variation could be used as a penalty function for the purpose of smoothness. 
Other alternative penalty functions include the label rank, 
Laplacian regularization and spectrum. 
In Section VI, we shall provide some filter design examples.

\section{Application Examples}
In this section, we consider several application examples for our newly 
proposed HGSP framework. These examples illustrate the practical use of HGSP 
in some traditional tasks, such as filter design and efficient data representation. 
We also consider problems in data analysis, such as classification and clustering.

\subsection{Data Compression}
Efficient representation of signals is important in data analysis and signal processing. Among 
many applications, data compression attracts significant interests for
efficient storage and transmission \cite{dc1,dc2,dc3}. 
Projecting signals into a suitable orthonormal basis is a widely-used compression method
\cite{c4}. Within the proposed HGSP framework, we propose a data compression method based on the hypergraph Fourier transform. 
We can represent $N$ signals in the original domain with $C$ frequency coefficients 
in the hypergraph spectrum domain. More specifically, with the help of the sampling theory 
in Section V, we can compress an $K$-bandlimited signal of $N$ signal points 
losslessly with $K$ spectrum coefficients.

\begin{table*}[htbp]
	\caption{Compression Ration of Different Methods} 
	\centering 
	\begin{tabular}{|l|l|l|l|l|l|l|l|l|l|l|l|}
		\hline
		size                                             & \multicolumn{7}{l|}{$16\times16$}                                & \multicolumn{3}{l|}{$256\times256$} &      \\ \hline
		image                                            & Radiation & People & load & inyang & stop & error & smile & lenna    & mri     & ct      & AVG  \\ \hline
		IANH-HGSP                                        & \textbf{1.52}      & \textbf{1.45}   & \textbf{1.42} & \textbf{1.47}   & \textbf{1.52} & \textbf{1.39}  & \textbf{1.40}  & \textbf{1.57}     & \textbf{1.53}    & \textbf{1.41}    & \textbf{1.47} \\ \hline
		($\alpha,\beta$)-GSP & 1.37      & 1.23   & 1.10 & 1.26   & 1.14 & 1.16  & 1.28  & 1.07     & 1.11    & 1.07    & 1.18 \\ \hline
		4 connected-GSP                                  & 1.01      & 1.02   & 1.01 & 1.01   & 1.04 & 1.02  & 1.07  & 1.04     & 1.05    & 1.07    & 1.03 \\ \hline
	\end{tabular}
	\label{t1}
\end{table*}

\begin{figure}[t]
	\centering
	\includegraphics[width=2.6in]{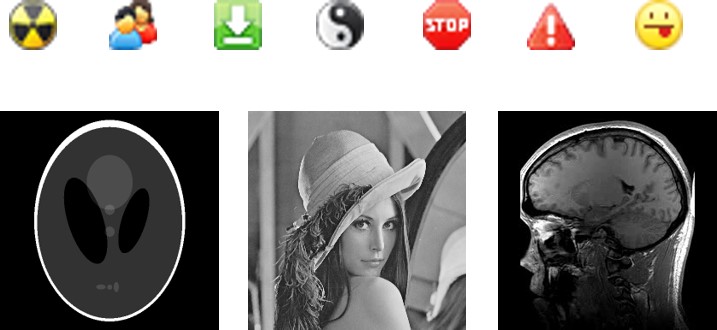}
	\caption{Test Set of Images.}
	\label{p1}
\end{figure}

To test the performance of our HGSP compression and 
demonstrate that hypergraphs may be a better representation of structured signals than 
normal graphs, 
we compare the results of image compression
with those from GSP-based compression method \cite{c4}.
We test over seven small size-$16\times16$ icon images and three size-$256\times256$ 
photo images, 
shown in Fig. \ref{p1}.

The HGSP-based image compression method is described as follows. Given an image, 
we first model it as a hypergraph with the Image Adaptive Neighborhood Hypergraph (IANH) model
\cite{c20}. To reduce complexity, we pick three closest neighbors 
in each hyperedge to construct a third-order adjacency tensor. Next, 
we can calculate the Fourier basis of the adjacency tensor as well as the bandwidth $K$ of 
the hypergraph signals. Finally, we can represent the original images using 
$C$ spectrum coefficients with $C=K$. For a large image, we may first cut it into smaller 
image blocks before applying HGSP compression to improve speed. 

For the GSP-based method in \cite{c4}, we represent the images as graphs with 1) the
4-connected neighbor model \cite{c21}, and 2) the distance-based model in which an edge exists only if
the spatial distance is below $\alpha$ and the pixel distance is below $\beta$. 
The graph Fourier space and corresponding coefficients in the frequency domain are then
calculated to represent the original image.

We use the compression ratio CR$=N/C$ to measure the efficiency of different
compression methods. A large CR implies higher compression efficiency. 
The result is summarized in Table \ref{t1}, from which 
we can see that our HGSP-based compression method achieves
higher efficiency than the GSP-based compression methods.

\begin{figure}[t]
	\centering
	\includegraphics[width=2.5in]{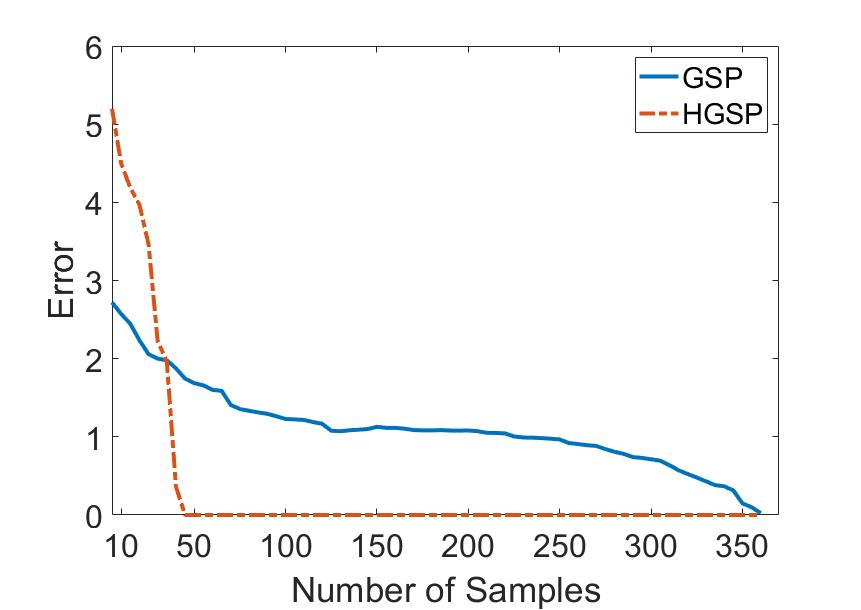}
	\caption{Errors of Compression Methods over the MovieLen Dataset: Y-axis shows the error between the original signals and recovered signals; X-axis is the number of samples.}
	\label{err}
\end{figure}

In addition to the image datasets, we also test the efficiency of HGSP spectrum compression 
over the MovieLens dataset \cite{data1}, where each movie data point has rating scores and tags from 
viewers. Here, we treat scores of movies as signals and construct graph models based 
on the tag relationships. Similar to the game dataset shown in Fig. \ref{hg2}, two movies 
are connected in a normal graph if they have similar tags. For example, if movies 
are labeled with `love' by users, they are connected by an edge. 
To model the dataset as a hypergraph, we include the movies into one hyperedge 
if they have similar tags. For convenience and complexity, 
we set $m.c.e=3$. 
With the graph and hypergraph models, we compress the signals using the sampling method
discussed earlier.  
For lossless compression, our HGSP method is able to use only $11.5\%$ of the samples from
the original signals to recover the original dataset by choosing suitable additional 
basis (see Section III-F). On the other hand, the GSP method requires $98.6\%$ of the samples. 
We also test the error between the recovered and original signals based on varying numbers of 
samples. As shown in Fig. \ref{err}, the recovery error naturally decreases with more samples. 
Note that our HGSP method achieves a much better performance once it obtains 
sufficient number of samples, while GSP error drops slowly. 
This is due to the first few key HGSP spectrum basis elements 
carry most of the original information, thereby leading to 
a more efficient representation for structured datasets.

Overall, hypergraph and HGSP lead to more efficient descriptions of structured data 
in most applications. With a more suitable hypergraph model and more developed methods, 
the HGSP framework could be a very new important tool in data compression.

\subsection{Spectral Clustering}
Clustering problem is widely used in a variety of applications, such as social network analysis, 
computer vision, and communication problems. Among many methods, spectral clustering is 
an efficient clustering method \cite{c26, c5}. Modeling the dataset by a normal 
graph before clustering the data spectrally,  significant improvement is possible
in structured data\cite{cc30}. However, such standard spectral clustering methods only
exploit pairwise interactions. For applications where the interactions 
involve more than two nodes, hypergraph spectral clustering should be a more natural choice.

In hypergraph spectral clustering, one of the most important issues is how to 
define a suitable spectral space. In \cite{c29,c30}, the authors
introduced the hypergraph similarity spectrum 
for spectral clustering. Before spectral clustering, they first modeled the 
hypergraph structure into a graph-like similarity matrix. They then defined the 
hypergraph spectrum based on the eigenspace of the similarity matrix. 
However, since the modeling of hypergraph with a similarity matrix 
may result in certain loss of the inherent information, a more efficient spectral space defined directly over hypergraph is more desired as introduced in our HGSP framework. 
With HGSP, as the hypergraph Fourier space from the adjacency tensor has a similar form 
to the spectral space from adjacency matrix in GSP, 
we could develop the spectral clustering method based on the hypergraph Fourier space as in
Algorithm 1.

\begin{algorithm}[htbp]
	\begin{algorithmic}[1] 
		\caption{HGSP Fourier Spectral Clustering}\label{basic3}
		\STATE {\bf{Input}}: Dataset modeled in hypergraph $\mathcal{H}$, the number of clusters $k$.
		\STATE Construct adjacency tensor $\mathbf{A}$ in Eq. (\ref{ad}) from the hypergraph $\mathcal{H}$.
		\STATE Apply orthogonal decomposition to $\mathbf{A}$ and compute
		Fourier basis $\mathbf{f_i}$ together with Fourier frequency coefficient $\lambda_i$ using Eq. (\ref{14}).
		\STATE Find the first $E$ leading Fourier basis $\mathbf{f}_i$ with $\lambda_i\neq 0$ and construct a Fourier spectrum matrix $\mathbf{S}\in\mathbb{R}^{N\times E}$ with columns as the leading Fourier basis.
		\STATE Cluster the rows of $\mathbf{S}$ into $k$ clusters using $k$-means clustering.
		\STATE Put node $i$ in partition $j$ if the $i$-th row is assigned to the $j$-th cluster.
		\STATE {\bf{Output}}: $k$ partitions of the hypergraph dataset.
	\end{algorithmic}
\end{algorithm}

To test the performance of the HGSP spectral clustering, we compare the achieved results with those from
the hypergraph similarity method (HSC) in \cite{c30}, using the zoo dataset\cite{c23}. 
To measure the performance, we compute the intra-cluster variance 
and the average Silhouette of nodes \cite{c31}. Since we expect the data points in the 
same cluster to be closer to each other, the performance is considered
better if the intra-cluster variance is smaller. 
On the other hand, the Silhouette value is a measure of how similar an object is to its own cluster versus
other clusters. 
A higher Silhouette value means that the clustering configuration is more appropriate. 

\begin{figure*}[t]
	\centering
	\subfigure[Variance in the same cluster]{
		\label{arbi}
		\includegraphics[height=5cm]{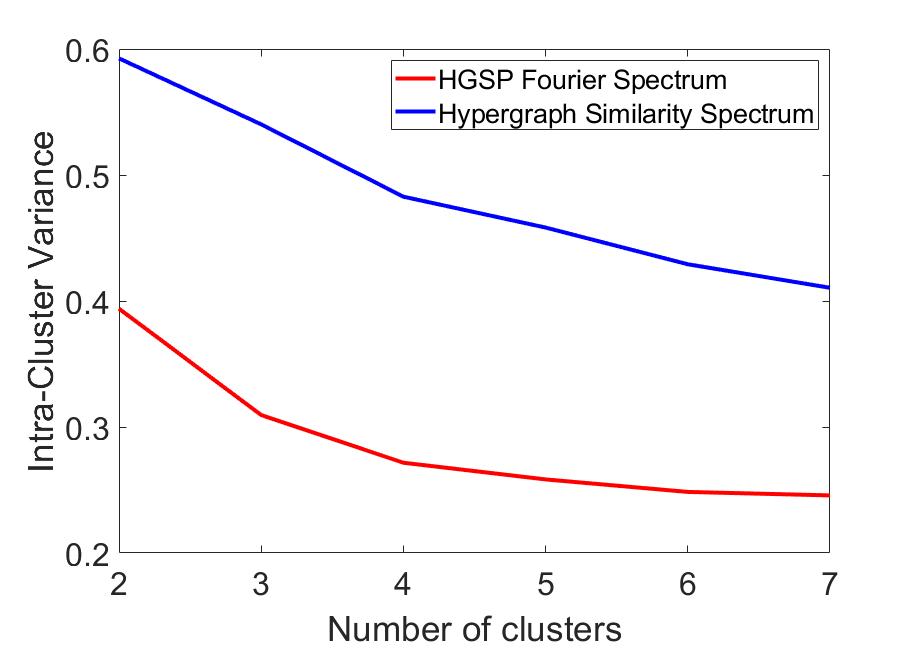}}
	\subfigure[Average Silhouette in the hypergraph]{
		\label{two}
		\includegraphics[height=5cm]{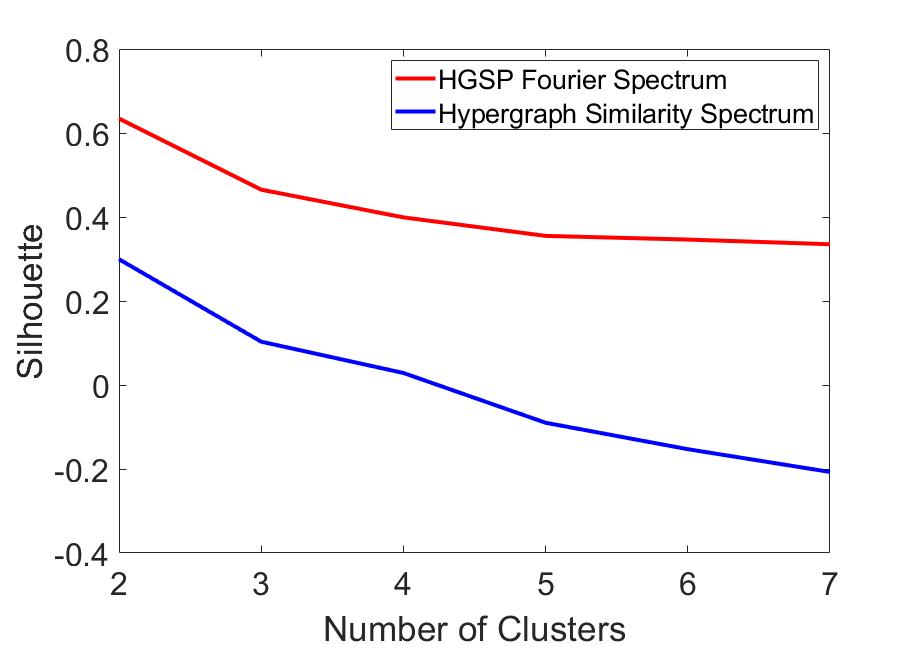}}
	\caption{Performance of Hypergraph Spectral Clustering.}
	\label{compare}
\end{figure*}

\begin{figure*}[t]
		\centering
	\includegraphics[width=6in]{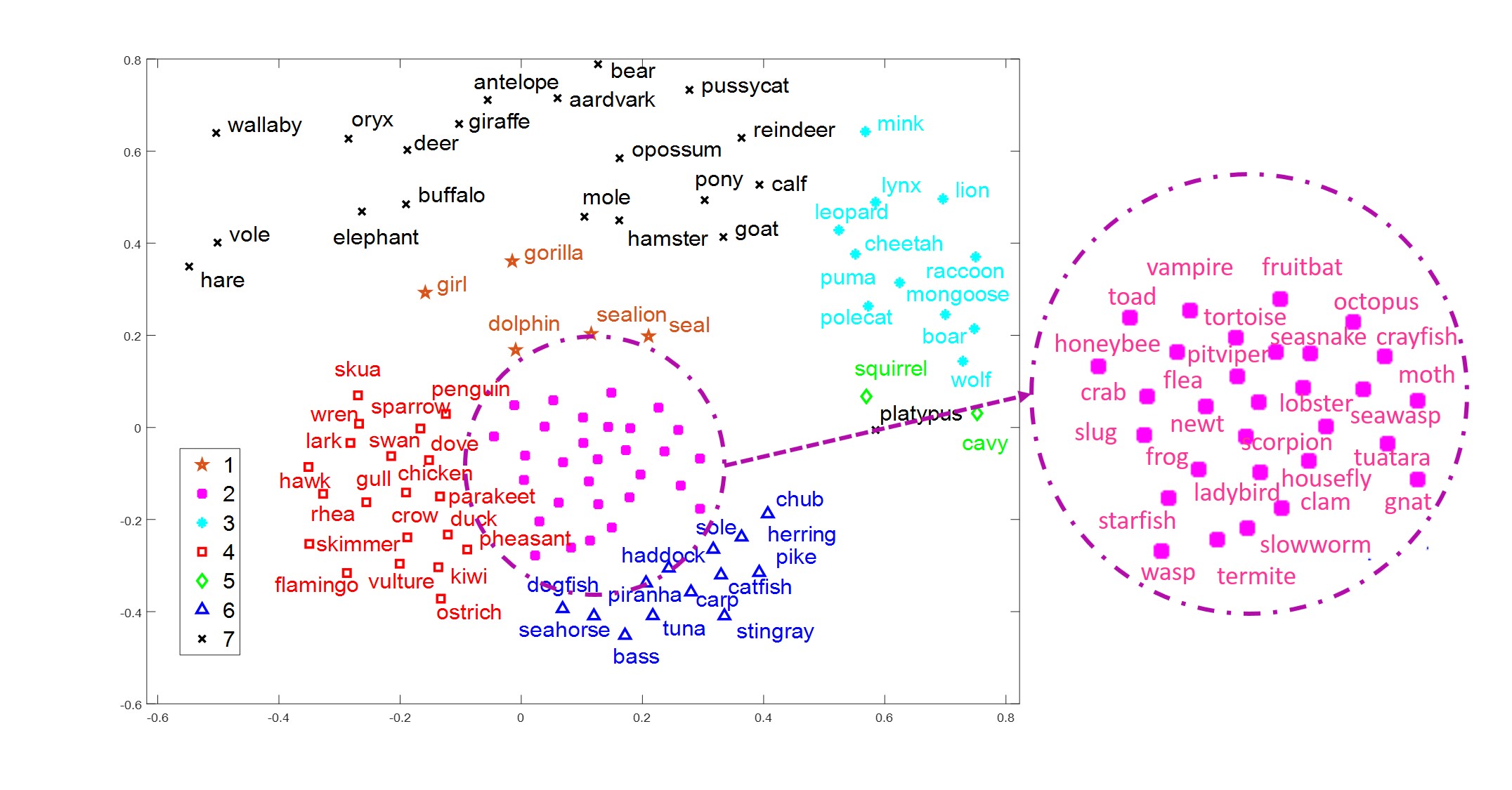}
	\caption{Cluster of Animals.}
	\label{clu1}
\end{figure*}

The comparative results are shown in Fig. \ref{compare}. 
Form the test result, we can see that our HGSP
method generates a lower variance and a higher Silhouette value.
More intuitively,  we plot the clusters of animals in Fig. \ref{clu1}. 
Cluster 2 covers small animals like bugs and snakes. 
Cluster 3 covers carnivores whereas cluster 7 groups herbivores. 
Cluster 4 covers birds and Cluster 6 covers fish. 
Cluster 5 contains the rodents such as mice. 
One interesting category is cluster 1: 
although dolphins, sea-lions, and seals live in the sea, 
they are mammals and are clustered separately from cluster 6.
From these results, we see that the HGSP spectral clustering method could achieve better 
performance and our definition of hypergraph spectrum may be
more appropriate for spectral clustering in practice. 


\subsection{Classification}
Classification problems are important in data analysis. Traditionally, these problems are studied 
by learning methods \cite{c24}. Here, we propose a HGSP-based method to solve the $\{\pm1\}$ classification problem, where a hypergraph filter serves as a classifier. 

The basic idea adopted for the classification filter design is label propagation (LP), where the main steps 
are to first construct a transmission matrix and then propagate the label based on the 
transmission matrix \cite{c25}. 
The label will converge after a sufficient number of shifting steps. 
Let $\mathbf{W}$ be the propagation matrix. Then
the label could be determined by the distribution $\mathbf{s'=W}^k\mathbf{s}$. 
We see that $\mathbf{s'}$ is in the form of filtered graph signal. 
Recall that in Section V-B, the supporting matrix $\mathbf{P}$ has been shown
to capture the properties of hypergraph shifting and total variation. 
Here, we propose a HGSP classifier based on the supporting matrix $\mathbf{P}$ 
defined in Eq. (\ref{P}) to generate matrix
\begin{equation}\label{LP}
\mathbf{H}=(\mathbf{I}+\alpha_1 \mathbf{P})(\mathbf{I}+\alpha_2 \mathbf{P})\cdots(\mathbf{I}+\alpha_k \mathbf{P}).
\end{equation}
Our HGSP classifier is to simply rely on
$
\mbox{sign}[\mathbf{Hs}].
$
The main steps of the propagated LP-HGSP classification method is described in Algorithm 2.
\begin{algorithm}[htbp]
	\begin{algorithmic}[1] 
		\caption{LP-HGSP Classification}\label{basic2}
		\STATE {\bf{Input}}: Dataset $\mathbf{s}$ consisting of labeled training data and unlabeled test data.
		\STATE Establish a hypergraph by similarities and set unlabeled data as $\{0\}$ in the signal $\mathbf{s}$.
		\STATE Train the coefficients $\alpha$ of the LP-HGSP filter in Eq. (\ref{LP}) by minimizing the errors of signs of training data in $\mathbf{s'=Hs}$.
		\STATE Implement LP-HGSP filter. If $\mathbf{s'}_i>0$, the $i$th data is labeled as $1$; otherwise, it is labeled as $-1$.
		\STATE {\bf{Output}}: Labels of test signals.
	\end{algorithmic}
\end{algorithm}

\begin{figure*}[t]
	\centering
		\centering
	\subfigure[Over datasets with a fixed datasize and different ratios of training data.]{
		\label{tdfd1}
		\includegraphics[height=5cm]{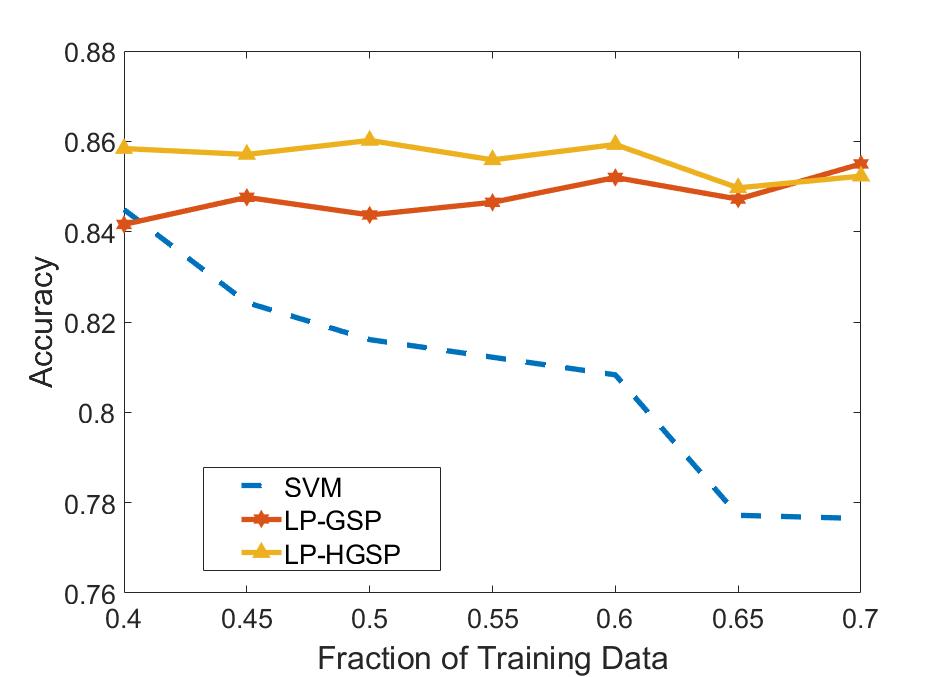}}
	\hspace{1cm}
	\subfigure[Over datasets with different datasizes and a fixed ratio of training data.]{
		\label{tdfd2}
		\includegraphics[height=5cm]{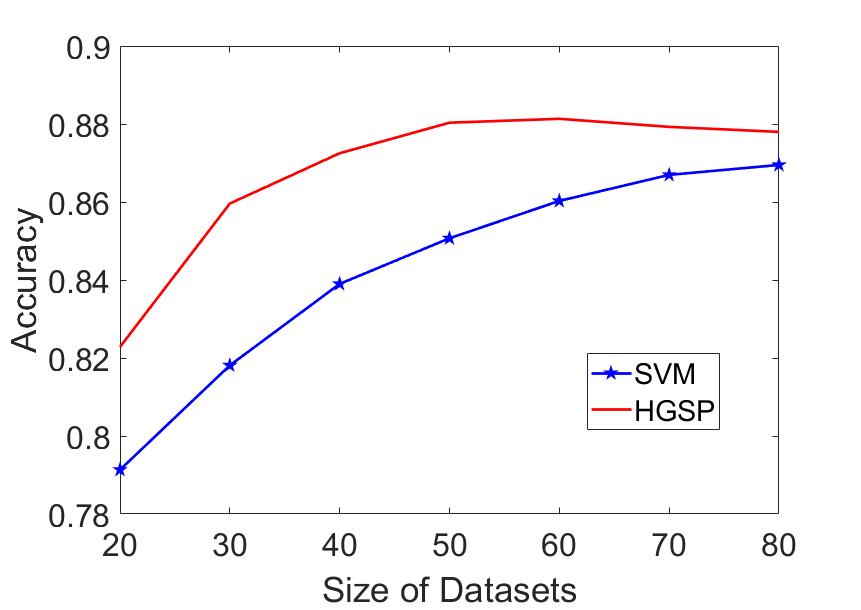}}	
	\caption{Performance of Classifiers.}
	\label{p2}
\end{figure*}

To test the performance of the hypergraph-based classifier, we implement them over the zoo datasets. We determine whether the animals have hair based on other features, formulated as a $\{\pm 1\}$ classification problem. 
We randomly pick different percentages of training data and leave the remaining data as the test set among the total 101 data points.
We smooth the curve with 1000 combinations of randomly picked training sets.
We compare the HGSP-based method against the SVM method with the RBF kernel and the label propagation GSP (LP-GSP) method \cite{c1}. 
In the experiment, 
we model the dataset as hypergraph or graph based on the distance of data. The threshold of determining the existence of edges is designed to ensure the absence of isolated nodes in the graph. 
For the label propagation method, we set $k=15$. The result is shown in Fig. \ref{tdfd1}. 
From the result, we see that the label propagation HGSP method (LP-HGSP) 
is moderately better than LP-GSP. 
The graph-based methods, i.e., LP-GSP and LP-HGSP, both perform 
better than SVM. The performance of SVM appears less satisfactory, likely because the dataset is rather small. 
Model-based graph and hypergraph methods are rather robust when applied to such small datasets. 
To illustrate this effect more clearly, we tested the SVM and hypergraph performance with
new configurations by the increasing dataset size and the fixing ratio of training data 
in Fig. \ref{tdfd2}. In the experiment, we first pick different sizes of data subsets from the original zoo dataset randomly as the new datasets. Then, with each size of the new dataset, $40\%$ data points are randomly picked as the training data, and the remaining data points are used as the test data. We average the results of 10000 times of experiments to smooth the curve.
We can see from Fig. \ref{tdfd2} that the performance of SVM shows 
significant improvement as the dataset size grows larger. 
This comparison indicates that SVM may require more data to achieve better performance, as shown in the comparative results of Fig. \ref{tdfd1}. Generally, the HGSP-based method exhibits better overall performance and
shows significant advantages with small datasets.
Although GSP and HGSP classifiers are both model-based, hypergraph-based ones usually perform 
better than graph-based ones, since hypergraphs provide a better description of the structured data in most applications.

\subsection{Denoising}
Signals collected in the real world often contain noises. 
Signal denoising is thus an important application in signal processing. 
Here, we design a hypergraph filter to implement signal denoising.

As mentioned in Section III, the smoothness of a graph signal, which describes the 
variance of hypergraph signals, could be measured by the total variation. Assume that the original 
signal 
is smooth. We formulate signal denoising as an optimization problem. 
Suppose that $\mathbf{y=s+n}$ is a noisy signal with noise $\mathbf{n}$, and $\mathbf{s}'=h(\mathbf{F,y})$ is the denoised data by the HGSP filter $h(\cdot)$. The denoising problem could be formulated as
an optimization problem: 
\begin{equation}\label{denoise}
\min_{h}||\mathbf{s'-y}||^2_2+\gamma \cdot ||\mathbf{s'}-\mathbf{s'}_{<1>}^{norm}||^2_2,
\end{equation} 
where the second term is the weighted quadratic total variation of the filtered signal $\mathbf{s'}$ based on the supporting matrix. 

The denoising problem of Eq. (\ref{denoise}) aims to smooth the signal 
based on the original noisy data $\mathbf{y}$. 
The first term keeps the denoised signal close to the original noisy signal, 
whereas the second term tries to smooth the recovered signal.
Clearly, the optimized solution of filter design is
\begin{equation}\label{d2}
\mathbf{s'}=h(\mathbf{F,y})=\mathbf{[I+\gamma(I-P_s)^{T}(I-P_s)]^{-1}y},
\end{equation}
where $\mathbf{P}_s=\sum_{i=1}^N \frac{\lambda_i}{\lambda_{\max}} 
\mathbf{f}_i \mathbf{f}_i^\mathbf{T}$ describes a hypergraph Fourier decomposition. 
From Eq. (\ref{d2}), we see that the solution is in the form of 
$\mathbf{s'=Hy}$ for denoising, which adopts a hypergraph filter $h(\cdot)$ as
\begin{equation}
\mathbf{H}=\mathbf{[I+\gamma(I-P_s)^{T}(I-P_s)]^{-1}}.
\end{equation} 

The HGSP-based filter follows a similar idea to
GSP-based denoising filter \cite{c22}. However, different definitions of the total 
variation and signal shifting result in different designs of 
HGSP vs. GSP filters. To test the performance, 
we compare our method with the basic Wiener filter, Median filter, and GSP-based filter \cite{c22} 
using the image datasets of Fig. \ref{p1}. We apply different types of noises. 
To quantify the filter performance, we use the mean square error (MSE) between 
each true signal and the corresponding signal after filtering.
The results are given in Table \ref{t11}. From these results, we can see that, 
for each type of noise and picking optimized $\gamma$ for all the methods, our HGSP-based filter out-performs other filters.

\begin{table*}[t]
	\caption{MSE of Filtered Signal} 
	\centering 
	\begin{tabular}{|l|l|l|l|l|l|l|l|}
		\hline
		$\gamma$ & 10e-5             & 10e-4     & 10e-3             & 10e-2      & 10e-1   & 1    & 10   \\ \hline
		\multicolumn{8}{|l|}{Uniform Distribution: U(0, 0.1)     }                                     \\ \hline
		GSP                   & 0.0031            & 0.0031   &0.0031 & 0.0026 & \textbf{0.0017} & 0.0895 & 0.4523 \\ \hline
		HGSP                  & 0.0031 & 0.0031 & 0.0028          & \textbf{0.0012} & 0.0631 & 0.1876 & 0.4083 \\ \hline
		Wiener &\multicolumn{7}{l|}{0.0201}              \\ \hline
		Median &\multicolumn{7}{l|}{0.0142}              \\ \hline
		\multicolumn{8}{|l|}{Normal Distribution: N(0, 0.09)      }                                      \\ \hline
		GSP                   & 0.790            & 0.790   & 0.0786 & \textbf{0.0556} & 0.0604 & 0.1286 & 0.4681 \\ \hline
		HGSP                  & 0.0790 & 0.0585 & \textbf{0.0305}          & 0.0778 & 0.1235 & 0.2374 & 0.4176 \\ \hline
		Wiener &\multicolumn{7}{l|}{0.0368}              \\ \hline
		Median &\multicolumn{7}{l|}{0.0359}              \\ \hline		
		\multicolumn{8}{|l|}{Normal Distribution: N(-0.02, 0.0001)      }                              \\ \hline
		GSP                   & \textbf{5.34e-04} & 5.36e-04 & 5.54e-04          & 7.76e-04 & 0.0055 & 0.1113 & 0.4650 \\ \hline
		HGSP                  & \textbf{4.17e-04} & 4.72e-04 &4.86e-04         &6.48e-04 & 0.0044 & 0.0868 & 0.3483 \\ \hline
		Wiener &\multicolumn{7}{l|}{0.0230}              \\ \hline
		Median &\multicolumn{7}{l|}{0.0096}              \\ \hline
	\end{tabular}
	\label{t11}
\end{table*}

\subsection{Other Potential Applications}
In addition to the application algorithms discussed above, there could be many other potential 
applications for HGSP. In this subsection, we suggest several potential applicable datasets and systems for HGSP.

\begin{itemize}
\item \textit{IoT:} With the development of IoT techniques, the system structures become increasingly complex, which makes traditional graph-based tools inefficient to handle the high-dimensional interactions. On the other hand, the hypergraph-based HGSP is powerful in dealing with high-dimensional analysis in the IoT system: for example, data intelligence over sensor networks, where hypergraph-based analysis has
already attracted significant attentions\cite{cc31}, and HGSP could be used to handle tasks like clustering, classification, and sampling. 
\item \textit{Social Network:} Another promising application is the analysis of
social network datasets. As discussed earlier, a hyperedge is an efficient representation for the multi-lateral relationship in social networks \cite{ota3,c9}; 
HGSP can then be effective in analyzing multi-lateral node interactions. 
\item \textit{Nature Language Processing:} Furthermore, natural language processing is an area that can benefit from
HGSP. Modeling the sentence and language by hypergraphs \cite{ota5,ota6}, HGSP 
can be a tool for language classification and clustering tasks. 

\end{itemize}
Overall, due to its systematic and structural approach, 
HGSP is expected to become an important tool 
in handling high-dimensional signal processing tasks that are traditionally addressed by DSP or GSP based methods.

\section{Conclusions}

In this work, we proposed a novel tensor-based framework 
of Hypergraph Signal Processing (HGSP) that generalizes the traditional GSP 
to high-dimensional hypergraphs. Our work provided important definitions in HGSP,
including hyerpgraph signals, hypergraph shifting, HGSP filters, frequency, and bandlimited signals.
We presented basic HGSP concepts such as the sampling theory and filtering design. 
We show that hypergraph can serve as an efficient model for many complex datasets. 
We also illustrate multiple practical applications for HGSP in signal processing and data analysis,
where we provided numerical results to validate the advantages and the practicality of 
the proposed HGSP framework. All the features of HGSP make it a powerful tool for IoT applications in the future.

\textit{Future Directions:} With the development of tensor algebra and hypergraph spectra, more opportunities are emerging to explore HGSP and its applications. One interesting topic is how to construct the hypergraph efficiently, where distance-based and model-based methods have achieved significant successes in specific areas, such as image processing \cite{hc2} and natural language processing \cite{hc3}. Another promising direction is to apply HGSP in analyzing and optimizing multi-layer networks. As we discussed in the introduction, hypergraph is an alternative model to present the multi-layer network \cite{mr3}, and HGSP becomes a useful tool when dealing with multi-layer structures. Other future directions include the development of fast operations such as the fast hypergraph Fourier transform, and applications over high-dimensional datasets \cite{vr}.


\ifCLASSOPTIONcaptionsoff
  \newpage
\fi

\end{document}